\documentclass[11pt]{article}
\usepackage{amsmath,amssymb}
\usepackage{epsfig}
\begin{document}
\title{A Two Stage Model for Quantitative PCR }
\author{
Emily Stone and John Goldes\\
{\it Dept. of Mathematical Sciences}\\
{\it The University of Montana}\\
{\it Missoula, MT 59801}
\\
and\\
Martha Garlick \\
{\it Department of Mathematics and Statistics}\\
{\it Utah State University}\\
{\it Logan, UT 84322-3900}}

\maketitle
\vspace{.25 in}

\begin{abstract}
 PCR (Polymerase Chain Reaction), a method which replicates a selected
sequence of DNA, has revolutionized the study of genomic material,
but mathematical study of the process has been limited to simple
deterministic models or descriptions relying on stochastic
processes. In this paper we develop a suite of deterministic models
for the reactions of quantitative PCR (Polymerase Chain Reaction)
based on the law of mass action. Maps are created from DNA copy
number in one cycle to the next, with ordinary differential
equations describing the evolution of difference molecular species
during each cycle. Qualitative analysis is preformed at each stage
and parameters are estimated by fitting each model to data from
Roche LightCycler (TM) runs.
\end{abstract}

\section{Introduction}
The Polymerase Chain Reaction (PCR), a technique for the enzymatic
amplification of specific target segments of DNA, has revolutionized
molecular biological approaches involving genomic material. This, in turn, has impacted
research in human genetics, disease
diagnosis, cancer detection, evolutionary and developmental
biology, and pathogen detection, to name a few.   The company Idaho Technology, Inc.
has capitalized on the invention
 of fluorescent probe techniques
 to create fast,
accurate devices for quantitative PCR.  Quantitative PCR is a method
where the amount of amplified DNA (or amplicon) is tracked
throughout the reaction and the initial amount of sample DNA can
then be estimated.   Understanding the important parts of a complex
reaction that is repeated tens of times, is critical in improving
the design of these processes in the laboratory, and to date
theoretical studies of quantitative PCR are limited.    In this
paper we present a suite of deterministic models for quantitative
PCR, with parameters estimated from data provided from Roche
LightCycler (TM) PCR runs.   Determining the critical features of
the model through  construction of increasingly complex descriptions
of the reaction is the overall goal of the project.

In PCR a reaction mixture containing a few copies of the target
double-stranded DNA is first heated to separate the DNA into single
strands.  It is then rapidly cooled and held at a lower temperature
briefly so that PCR primers (short single  strands of DNA that have
been designed for this purpose) anneal specifically to the template
DNA.  The enzyme Taq Polymerase recognizes these  primer-template
pairs and synthesizes a new strand of DNA, starting at the end of
the annealed primer.  In this way, a complementary strand is made
from each strand of the original double-stranded DNA molecule.
Under ideal reaction conditions the number of copies of this stretch
of DNA in the sample is doubled in each heating-cooling cycle.

Instruments that perform real-time PCR usually detect the amplified DNA
using fluorescent probes that are added to the PCR reagents before
temperature  cycling.  These probes bind to the DNA and generally
fluoresce more when bound than when free.  When there is a sufficient
quantity of DNA present
in the sample (for example, after many temperature cycles), this change
in fluorescence is detected using a fluorimeter.  If the fluorescent
signal of a sample rises above a background level, a sizable amount of
DNA has been synthesized, indicating that the specific DNA was initially
present.

Current methods for DNA quantification (for more information see the
following references: Morrison et al. (1998), Wittwer et al. (1994,
1997), Weiss and Von Haeseler (1997), Sun (1995), Sun et al. (1996))
with PCR are based on comparing a set of successively diluted
standards against unknown samples.  The methods utilize the
concentrations of the standards in a dilution series to determine
the concentration of the unknown.  The amount of DNA in successively
diluted standards is typically decreased by factors of 2 or 10, and
anywhere from three to ten standards are used.  Figure 1 shows a set
of six standards containing between one and 1,000,000 copies of
initial DNA template.   The fluorescence curve that crosses the
threshold value at the smallest cycle number initially had 1,000,000
copies of DNA, the next curve to cross the threshold had 100,000
copies of DNA initially, and so on.  Notice that the curve that does
not cross the threshold is the control-no-template sample. Also
plotted on this graph are two sets of replicates with unknown
initial quantities of DNA template.

\begin{figure}
\centerline{\epsfig{file=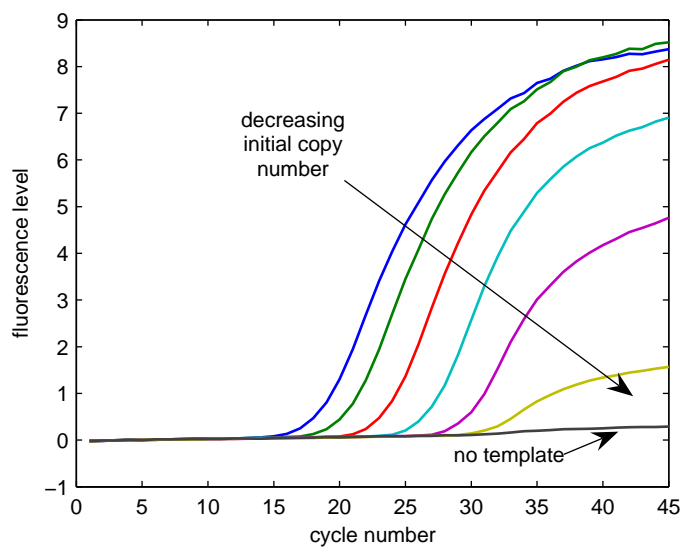}}
\caption{Fluorescence level vs. cycle number during PCR Roche Lightcycler run.
Different lines are standard dilutions for quantification purposes, from
$10^6$ copies down to $10^1$ and no template as a control. Also run simultaneously
are two samples of unknown concentration, five replicates each.}
\label{fluor}
\end{figure}

A quick estimate of the order of magnitude of the number of copies of
DNA initially in the unknown samples can be found by simply comparing the
amplification curves of the samples and diluted standards.  Current
methods produce more precise estimates using a
mathematical model of PCR that assumes the product grows exponentially:

\begin{equation}
C_{n+1} = C_{n}  + E C_{n} =  (1+E) C_{n}= (1+E)^n C_0.
\label{expmodel}
\end{equation}

In this model, $C$ represents the number of copies of DNA, $n$ represents
the cycle number and $E$ represents the efficiency of the PCR.  $E$ can be
thought of as the percentage of existing DNA that is replicated in a
cycle.  Dilution standards have known values for $C_0$, and data from these
samples can be used to calculate the efficiency $E$.  Given the
efficiency, the initial copy number $C_0$ can be estimated for each unknown
sample.

This model is accurate for a small number of cycles, but grows less and
less accurate as the number of DNA copies grows.  Unfortunately, the
fluorescence signal can be distinguished from the noise only later in
the experiment, precisely as the model becomes less accurate.  The
simplest mistake in the model is the assumption that the efficiency
does not change with cycle number, and that the number of copies of DNA
always grows.  In reality, PCR products saturate the reaction and
resources are exhausted, slowing and eventually stopping DNA synthesis.

The amplification curves thus suggest that a more natural model for
the PCR reaction would be logistic, which proceeds to saturation as
a resource is depleted.   Current IT software fits the data to such
a logistic map, and uses the result to estimate initial copy number
of the template.  For the purpose of this estimation both the
exponential growth model and the logistic model are sufficient in
many cases, and have the advantage of a limited number of free
parameters, requiring a minimum amount parameter estimation.
However, for the long range goal of developing a more complete model
of the reaction that can lead to innovation in process design, we
must look beyond these one dimensional approximations. We also see
that the data deviate from the logistic model in a consistent way
for all the amplification curves, suggesting that the
simplifications leading to it eliminate some critical features of
the dynamics.

To our knowledge no deterministic model of the reactions
of PCR that does not include assumptions about the kind of enzyme
kinetics involved (i.e. Michaelis-Menten) are present in the
literature.   Stochastic models for estimating reaction efficiency
and specificity can be found however, for instance, in
Sun (1995) a
model for distributions of mutations and estimation of mutation
rates during PCR is developed,
using the theory of branching processes.
Another such model is reported in Weiss and Von Haeseler (1995), where the accumulation
of new molecules during PCR is treated as a randomly bifurcating tree
to estimate overall error rates for the reaction.
In Schnell and Mendoze (1997), the reaction efficiency of quantitative competitive PCR (QC-PCR:
a target and a competitor template are amplified simultaneously
to provide an internal standard for identifying the initial target template amount)
is computed using Michaelis-Menten type kinetics.

Stolvitzky and Cecchi (1996)
address the validity of assuming a constant efficiency during a PCR
reaction by deriving the probability of replication during successive
cycles as a function of physical parameters.   In the same vein,
Velikanov and Kapral (1999) report on a probabilistic model
of the kinetics of PCR using microscopic Markov processes.   The result
is an exact solution for the distribution of lengths of synthesized
DNA strands, and an optimization procedure is applied to determine
control parameters that maximized the yield of the target sequence.
Most recently, in a 2004 publication Whitney et al. describe a
stochastic model for competitive interactions during PCR to compute product distributions
at the completion of regular PCR.  The calculated yield is compared to
experimental values from the amplification of three different size amplicons,
with good results.

In this paper we develop deterministic models based directly on the
reaction equations using the law of mass action. A hierarchy of
models is built by including more biochemistry into each successive
level of approximation. We analyze qualitatively and numerically the
solutions to the models under typical operating conditions, and
perform parameter estimation  with data provided by Idaho
Technology. Finally the advantages and disadvantages of including
more details of the reactions into the model are discussed.

\section{The Reactions of PCR}

\begin{center}
Table I: List of Variables and Notation Used in PCR Models
\begin{tabular}{|c|l|}      \hline
Variable name   &  quantity \\ \hline
$C$ & copy number \\
$E$ & exponential efficiency of reaction\\
$S, s$  &   single stranded DNA (ssDNA), $s=[S]$\\
$P, p$  &   primer molecule, $p=[P]$\\
$S', s'$ &   primed ssDNA, $s'=[S']$\\
$Q, q$  &   Taq molecule,  $q=[Q]$\\
$C, c$  &   enzyme complex, $c=[C]$\\
$N, n$  &   nucleotide sequence for the extension, $n=[N]$\\
$D, d$  &   double stranded DNA (dsDNA), $d=[D]$\\
$k_{-1},k_{1}$ &forward and backward reaction rates for annealing \\
$k_{-2},k_{2}$ &forward and backward reaction rates for complex formation \\
$k_{-3},k_{3}$ &forward and backward reaction rates for extension \\
$ \epsilon$  & logistic map growth parameter \\
$ \delta$  & logistic model growth parameter \\
$\cal{K}$ & carrying capacity of the logistic map\\
$\Gamma(d_i)$ & growth parameter function for Taq model\\
$e, \alpha$ & parameters in $\Gamma(d_i)$ \\
$W_i$  &  estimation of $\Gamma(d_i)$ from experimental data\\
$Y_i$  &  logarithmic regression variable\\
$\Delta t$ & time step in discrete version of logistic model\\
$t_1$  & time in stage I of two stage model without Taq dynamics\\
$t_2$  & time in stage II of two stage model without Taq dynamics\\
$\tau_I$  & rescaled time in stage I of two stage model without Taq dynamics\\
$\tau_{II}$  & rescaled time in stage II of two stage model without Taq dynamics\\
$K, K_n, K_d, K_s$ & conserved quantities in the two stage models\\
$KK $  & normalization parameter used for experimental data\\
$\beta, \gamma $ & rescaled reaction rates in the two stage model with Taq dynamics\\
$s'^I,s'^{II}$ & $s'$ in stage I and stage II respectively\\
$\bar{x}$  & fixed point of the $x$ variable\\
$t_I,t_{II}$ & time in stage I and stage II in model with Taq dynamics\\

\hline

\end{tabular}
\end{center}

The PCR reaction proceeds through repeated cycles of dissociation,
annealing and extension by the enzyme Taq polymerase.  During
dissociation the sample is heated to approximately 90 degrees C
where the template's DNA nucleotide base pairs unbind and the strand
essentially unzips to form two half-strands  (single stranded DNA).
The sample is then cooled to a temperature where the primer reaction
is optimal (about 60 degrees C), during which primer molecules,
themselves sequences of single stranded DNA that have been designed
to adhere to either end of the target sequence of the template, bind
on.  Then the sample is heated again to a temperature where Taq
enzyme adds base pairs on the bracketed sequence to form a new
double-stranded piece of DNA. The annealing/extension can done in
one or two distinct steps, either with a continuous ramp-up to the
Taq operating temperature (during which time the primers anneal) or
with a lower temperature annealing stage followed by a higher
temperature extension phase.   We model the latter, but the model
itself could easily be adapted for the one-step scenario.

These three phases, dissociation, annealing, and extension, are
repeated typically 30-40 times to yield exponentially growing
numbers of the target sequence, assuming the reaction runs as
designed.  Factors influencing the success of the reaction (is there
a product?) are competition from contaminants in the reaction
mixture, primers that bind to themselves or other primer molecules
(primer-dimers), or primers that can extend pieces of the template
other than the target, to name a few. Naturally the reaction
saturates, see fig. \ref{fluor}, which is assumed to occur by
complete depletion of primer molecules, since they are incorporated
into the extended strands. The nucleotides in the mixture could also
be used up, but typically they are present in great numbers to
prevent this from occurring.  Primers are synthesized molecules and
are therefore much more costly than nucleotides.  Also,  the initial
amount of DNA to be amplified can not be either too large or too
small.  If it is too large the number of primers is not sufficient
to completely prime the molecules, and if too small it can lose out
to the competing amplification of undesired sequences.

The reaction equations for these phases can be written as follows.\\
Dissociation:
\[
D \rightarrow  2 S
\]
Annealing:
\[
S+P \begin{array}{c} \underrightarrow{\scriptstyle{k_{1}}}\\
\overleftarrow{\scriptstyle{k_{-1}}} \end{array} S'
\]
Extension:
\[
S'+Q \begin{array}{c} \underrightarrow{\scriptstyle{k_{2}}}\\
\overleftarrow{\scriptstyle{k_{-2}}} \end{array} C
\]
\[
C+N \begin{array}{c} \underrightarrow{\scriptstyle{k_{3}}}\\
\overleftarrow{\scriptstyle{k_{-3}}} \end{array} D + Q
\]
Here $D$ is double-stranded DNA, $S$ is single-stranded DNA, $P$ is
primer, $S'$, primed single-stranded DNA, $Q$, Taq polymerase, $C$,
complex of primed single-stranded DNA, $P'$, and Taq, $N$,
nucleotides.  The plus/minus $k$'s represent the forward and
backward reaction rate respectively.  Ideally the reactions form a
cascade, the product of one reaction continues into the next
reaction and the final double-stranded DNA cycles back to the
dissociation phase.  In reality the reactions occur simultaneously,
with highest frequency at their optimal temperature.  For our
purpose we will treat the phases as distinct and cascade the output
of one phase to the input of the next. We also assume that the back
reactions are negligible compared to the forward reactions in all
but the creation of the enzyme complex, e.g. $k_{-1}=k_{-3}=0$.

The law of mass action can be invoked to create differential equations
for the concentrations of the above reactants, and we use lower case
letters to indicate these concentrations, e.g. $[S]=s$, $[D]=d$, etc.
We assume that the resource, nucleotides, is present in chunks of
appropriate sequences of base pairs for the segment of DNA being extended.
That is, we will assume that the extension happens all at once, not
one base pair at  a time.
For the annealing reaction we have:
\begin{equation}
\frac{ds}{dt}=-k_1 s p
\label{a1}
\end{equation}
\begin{equation}
\frac{dp}{dt}=-k_1 s p
\label{a2}
\end{equation}
\begin{equation}
\frac{d s'}{dt}=+k_1 s p.
\label{a3}
\end{equation}
And for the extension phase the equations are:
\begin{equation}
 \frac{d s'}{dt}=-k_2 s' q +k_{-2} c
\label{e1}
\end{equation}
\begin{equation}
\frac{dq}{dt}=-k_2 s' q+k_{-2} c+ k_3 c n
\label{e2}
\end{equation}
\begin{equation}
\frac{dc}{dt}=k_2 s' q-k_{-2} c-k_3 c n
\label{e3}
\end{equation}
\begin{equation}
\frac{dn}{dt}=-k_3 c n
\label{e4}
\end{equation}
\begin{equation}
\frac{dd}{dt}=k_3 c n.
\label{e5}
\end{equation}

The exponential model (\ref{expmodel}) is a first level of
approximation to the growth of double-stranded DNA created in these
reactions, and the next level of simplification is the logistic map.
This can be arrived at in a  straightforward manner from these
reaction equations by first ignoring the enzyme (Taq) dynamics.
Working with
the set of equations that result if the Taq dynamics is ignored we have:\\
Annealing (as above):
\[
\frac{ds}{dt}=-k_1 s p
\]
\[
\frac{dp}{dt}=-k_1 s p
\]
\[
\frac{d s'}{dt}=+k_1 s p
\]

Extension:
\begin{equation}
\frac{d s'}{dt}= -k_3 s' n
\label{ee1}
\end{equation}
\begin{equation}
\frac{dn}{dt}= -k_3 s' n
\label{ee2}
\end{equation}
\begin{equation}
\frac{dd}{dt}= k_3 s' n
\label{ee3}
\end{equation}

If the annealing stage is assumed to achieve 100\% priming, the
output of the first three equations is $s'(t_{end})=s(t_0)$, which
in turn becomes the initial condition into the extension phase.
Using the conserved quantity $d+n$ in the extension phase, writing
$d+n={\cal K}=d(0)+n(0)$ and setting $n={\cal K}-d$ makes the
equation for $d$:
\[
\frac{dd}{dt}=k_3 s' ({\cal K}-d).
\]
Taking an Euler step approximation to this ODE yields:
\[
\Delta d = k_3 \Delta t s' ({\cal{K}}-d).
\]
If the time step $\Delta t$ is taken to be the time for one cycle,
the result is a map for amounts of $s'$, $d$, and $n$ from one cycle
to the next. For the $i$-th cycle:
 \[
\Delta d = d_{i+1}-d_{i}= \epsilon s'_i ({\cal{K}}-d_i).
\]
where $\epsilon = \Delta t k_3$. With perfect priming $s'_i = s_i$,
the amount of single stranded DNA at the beginning of the priming
phase, and with perfect dissociation $s_i=2 d_i$, the amount of
double stranded DNA from the end of last extension phase. The
equation for $d$ then becomes
\begin{equation}
d_{i+1}= d_{i} + \epsilon 2 d_i ({\cal{K}}- d_i), \label{logmodel}
\end{equation}
which is a logistic map for $d_i$.

To test the assumption of logistic data, and we fit a dilution
series from a Roche LightCycler (TM) run to (\ref{logmodel}).   The
run was typical for these quantification experiments: it had 45
cycles, each consisting of a brief melt stage at 95 degrees C, a 10
second annealing stage at 55 degrees C,  and a 30 second extension
stage at 72 degrees C. The fluorescence acquisition occurred at the
end of the annealing stage, and used a FRET (fluorescence resonance
energy transfer) probe system. FRET probes are a pair of
oglionucleotides labeled with fluorescent dyes. The pair are
designed to hybridize to adjacent regions on the target DNA, and the
marker dyes of each probe can only interact when they are in close
proximity and bound to the target.   The fluorophores are chosen so
that the emission spectrum of one overlaps with the excitation
spectrum of the other. The donor fluorophore is excited by a light
source, transfers its energy to an acceptor fluorophore, which then
emits light of a longer wavelength. This light is then detected
during the fluorescence acquisition.

The parameter estimation was done in MATLAB using least squares to
first compute $K$ and $\gamma$, and a simplex search method based
scheme (MATLAB's {\it fminsearch}) to find the initial fluorescence
level. The objective function used in the nonlinear optimization was
the two-norm of the difference between the model time series and the
data. Percentage error was computed by dividing the final value of
the objective function by the two-norm of the model time series.

\begin{center}
Table II: Parameter Estimation for the Logistic Model (\ref{logmodel})

\begin{tabular}{|l|c|c|c|c|c|}      \hline
    & $\gamma$  & $K$       &$d_0$ & \% error\\ \hline
run 1   & 0.0185  &8.1473   &   $2.9\times10^{-2}$ & 5.15 \\  \hline
run 2   & 0.0189  &8.2832   &   $1.6\times10^{-2}$ & 5.8\\  \hline
run 3   & 0.0214  & 7.9001&     $4.2\times10^{-3}$ & 5.15 \\\hline
run 4   & 0.0296  & 6.678   &       $5.5\times10^{-4}$ & 4.9 \\\hline
run 5   & 0.0457 &  4.5959  &   $1.02\times10^{-5}$ & 5.3\\  \hline
run 6   & 0.1273   & 1.5445 &   $4.12\times10^{-5}$ & 5.82\\  \hline
\end{tabular}
\end{center}

\begin{figure}
\centerline{\epsfig{file=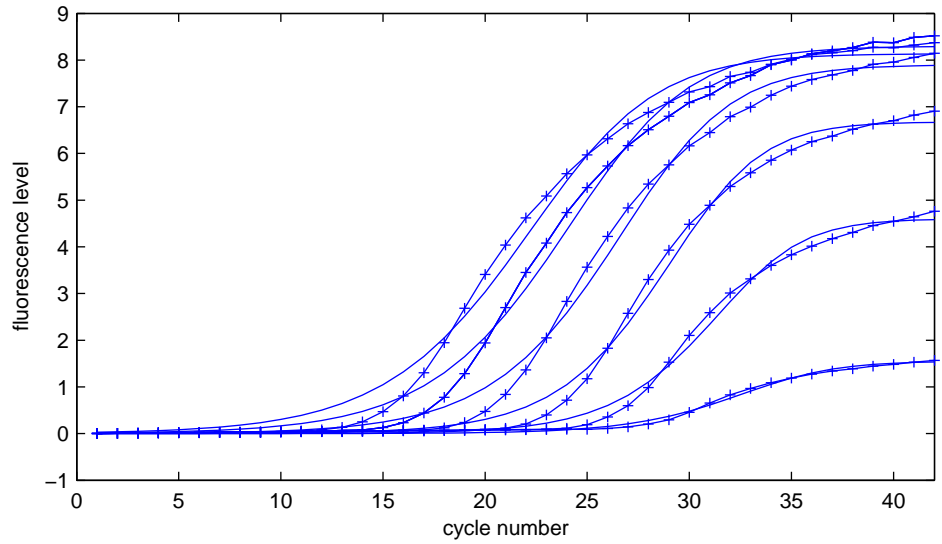}}
\caption{Fitting dilution series data with a logistic map.  See text for
parameter values.  Solid line- model, (+++++)- data.}
\label{LSQfit}
\end{figure}

The results of the parameter estimation for the dilution series are
present in figure \ref{LSQfit} and Table II.    We see that the
model is more than adequate for predicting initial copy number,
given the current practice of running standards simultaneously with
samples to generate a map between initial copy number and
fluorescence level.    The drift in the growth constant for
decreasing copy number indicates that some aspect of the dynamics is
not captured by this map, and for the lowest copy number run (number
6) we see that the initial amount estimated is off by an order of
magnitude.   In this case competition from other reactions is
thought to be the culprit, but in all cases the map is a gross
simplification to the actual dynamics.  It clearly overestimates the
growth for earlier cycles, and approaches saturation more quickly at
later cycles. To verify this intuition quantitatively we preformed a
version of logarithmic regression on the first five runs of the
dilutions series (see figure \ref{fluor}). The basis for this
regression is the separation of variables solution to the logistic
differential equation:
\[
\frac{dy}{dt}=\delta y (K-y),
\]
namely
\[
\ln(\frac{y}{K-y})=K \delta t + \ln(\frac{y_0}{K-y_0}).
\]

\begin{figure}
\centerline{\epsfig{file=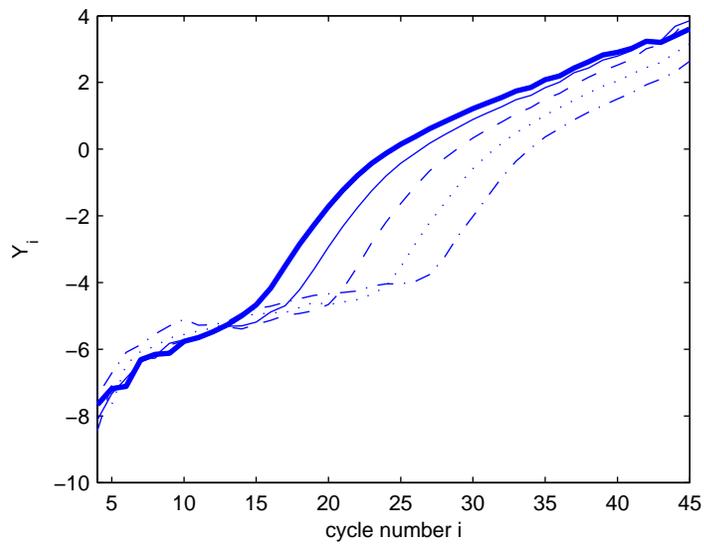}}
\caption{Logarithmic regression curves for five standard dilution series.}\label{logreg}
\end{figure}

\begin{figure}
\centerline{\epsfig{file=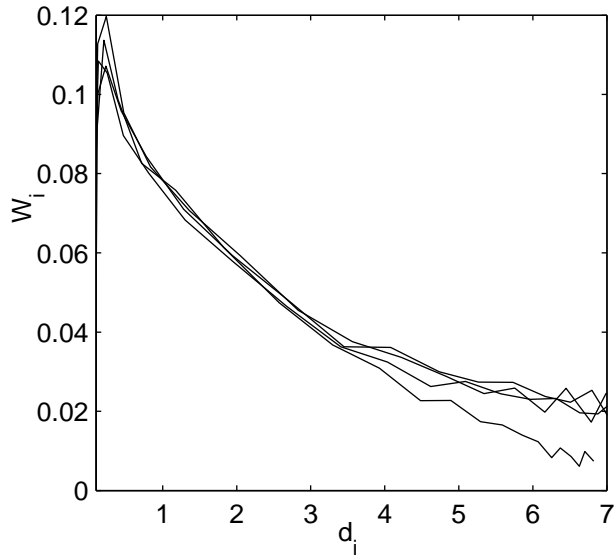}} \caption{Plotting $W_i$ vs.
$d_i$ to estimate $\Gamma(d_i)$ for the first four dilution series
runs.}\label{taqfunc}
\end{figure}

\begin{figure}
\centerline{\epsfig{file=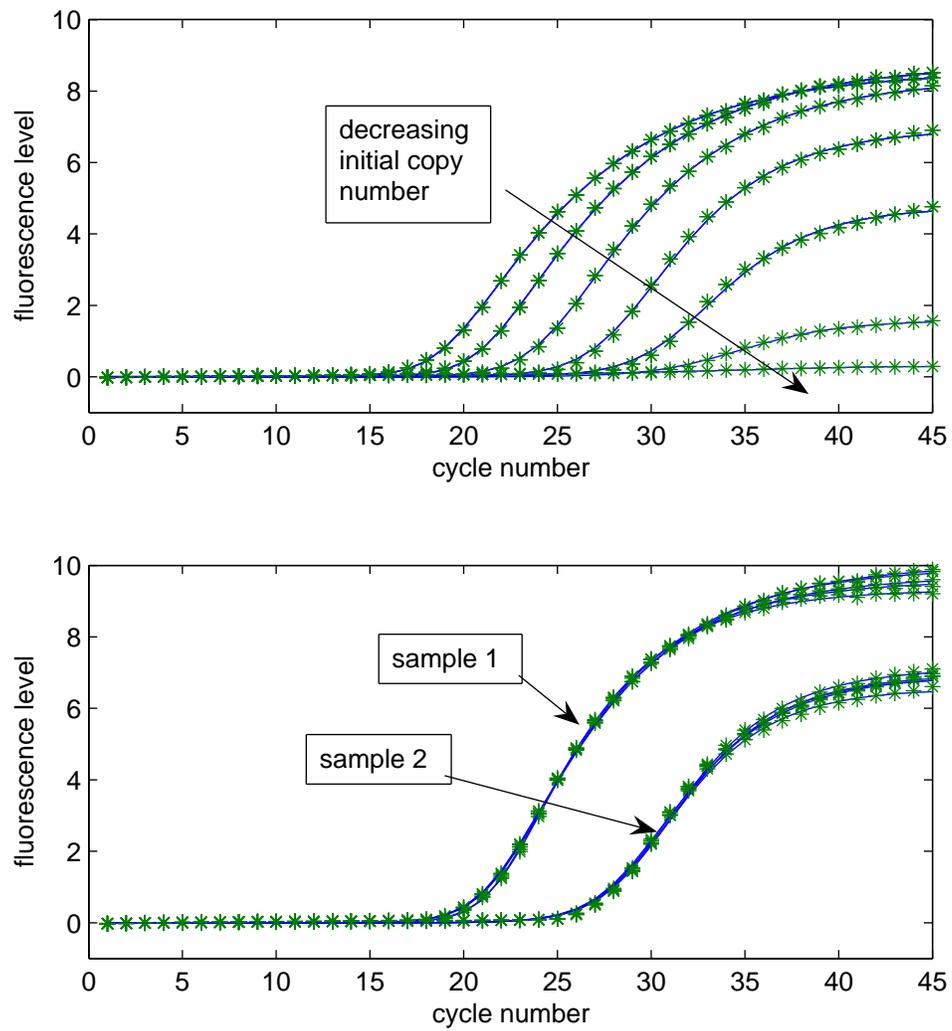}}
\caption{Standards data fitted with the Taq model (\ref{Taqmodel}). See Table II for
parameter and error information.}\label{fittaqmodel}
\end{figure}

If $y$ follows logarithmic growth the variable $Y =
\ln(\frac{y}{K-y})$ will depend linearly on time $t$.   Plots of the
discretely sampled variable $Y_i$ vs. cycle number $i$ for standard
data set are shown in figure \ref{logreg}, where it is obvious that
they are not well-estimated by a linear function of $i$. There are
several straight line regions in these graphs, corresponding to a)
low cycle number noise, b) a region where the initial exponential
growth occurs, and c) a third region where saturation happens, with
a slope less than that seen in the second region.  A linear fit of
this data would have an intermediate slope causing an overestimate
or an underestimate of the data, depending on the cycle number. This
is apparent in figure \ref{LSQfit}, where it also is clear that the
data approach the saturation level more slowly than logistic growth
would warrant. One explanation of this is reduced efficiency of Taq
polymerase when the quantity of molecules to extend becomes very
large. This suggests a growth parameter $\Gamma(d_i)$ that varies
with the amount of amplicon, $d$, so that
\[
d_{i+1}= d_{i} + \Gamma(d_i) d_i (K- d_i),
\]
where $\Gamma (d_i)  $ is a decreasing function of $d_i$.  The shape of this function
can be estimated by plotting the variable found by solving the above
equation for $\Gamma$:
\[
W_i=\frac{d_{i+1}-d_i}{d_i (K- d_i)}= \Gamma(d_i).
\]
That is, the graph of  $W_i$ vs. $d_i$ will give an idea of $\Gamma(d_i)$.   An
example is shown in figure \ref{taqfunc}.   The function $\Gamma$ appears to
be inversely proportional to $d_i$, so we fit a new map with
$\Gamma(d_i)=  \frac{e}{1+\alpha d_i}$,
so
\begin{equation}
d_{i+1}= d_{i} +  \frac{e d_i}{1+\alpha d_i} (K- d_i).
\label{Taqmodel}
\end{equation}
The results are shown in figure \ref{fittaqmodel}, and in Table III.
The error presented there is the mean square error.

\eject

\begin{center}
Table III: Parameter Estimation of Taq Model (\ref{Taqmodel}).
\begin{tabular}{|l|c|c|c|c|c|}      \hline
    &error  &$e$    &$K$    &$\alpha$   &$d_0$\\ \hline
run 1   &0.1894 &0.1481 &8.5579 &0.9148 &$4.91\times10^{-7}$\\  \hline
run 2   &0.2055 &0.1473 &8.7559 &0.8907 &$7.31\times10^{-8}$\\  \hline
run 3   &0.2587 &0.1227 &8.3295 &0.6096 &$1.20\times10^{-7}$\\  \hline
run 4   &0.3639 &0.1237 &6.9411 &0.4158 &$1.06\times10^{-7}$\\  \hline
run 5   &0.3945 &0.1592 &4.7559 &0.4111 &$7.53\times10^{-8}$\\  \hline
run 6   &0.2394 &0.3721 &1.5740 &0.6636 &$3.02\times10^{-7}$\\  \hline
sample 1a &0.2321 &0.1413 &9.3277 &0.5918 &$3.98\times10^{-8}$\\  \hline
sample 1b &0.2377 &0.1546 &9.5827 &0.7381 &$1.04\times10^{-8}$\\  \hline
sample 1c &0.2189 &0.1231 &9.7042 &0.5957 &$1.57\times10^{-7}$\\  \hline
sample 1d &0.2582 &0.1202 &9.9448 &0.6035 &$1.60\times10^{-7}$\\  \hline
sample 1e &0.2610 &0.1196 &10.0509&0.6391 &$1.62\times10^{-7}$\\  \hline
sample 2a &0.4427 &0.1297 &6.5511 &0.3543 &$8.28\times10^{-8}$\\  \hline
sample 2b &0.4203 &0.1222 &6.8775 &0.3390 &$9.03\times10^{-8}$\\  \hline
sample 2c &0.4328 &0.1214 &6.9499 &0.3581 &$9.08\times10^{-8}$\\  \hline
sample 2d &0.4223 &0.1173 &7.1260 &0.3541 &$1.07\times10^{-8}$\\  \hline
sample 2e &0.4270 &0.1267 &6.9205 &0.3514 &$6.42\times10^{-8}$\\  \hline
\end{tabular}
\end{center}

The growth coefficient for this model ($e$), and the value for
$\alpha$ are more consistent than for the regular logistic model,
though the variation is more pronounced in smaller copy number runs,
and suffers the same overcalculation of the initial fluorescence in
run 6.  (Also in run 6 the model coefficients are significantly
different from the other runs).   We also fitted the replicates of
the unknown samples (sample 1 and sample 2) with reasonably
consistent results, though it is clear there is a trade-off between
values such as the growth constant $e$ and the initial fluorescence,
indicating hidden dependencies in the parameters that cannot be
differentiated with this sort of data.

This could be the end of the story, but an empirically determined
rate function is not as satisfactory as a model that captures the
behavior built-up directly from the reaction equations. In the next
section we construct such a model and parameterize it with the data.

\section{The Two Stage Model}
We consider here two versions of a two stage model, one that includes
the dynamics of the Taq enzyme, and one that does not.  The latter is
much simpler and can be solved analytically, so we present it first.
Both make the assumption of complete dissociation, so that the amount
of ssDNA entering the annealing phase is equal to twice the amount of
dsDNA from the previous extension phase, plus whatever ssDNA was leftover
from the previous annealing phase.
This eliminates the need for the equation that describes dissociation (or
the ``melt") phase of the reaction.  We next
assume that the annealing phase happens distinct from the extension phase and
call this stage I.  The equations for stage I are   (\ref{a1})-(\ref{a3}).
The extension phase we name stage II, and with Taq dynamics the equations
are (\ref{e1})-(\ref{e5}).   Without the Taq dynamics the equations are
given by (\ref{ee1})-(\ref{ee3}).

\subsection{Two Stage Model without Taq Dynamics}
The two stages in both versions are linked through their initial conditions,
and in both the initial amount of primed ssDNA in stage II is equal to the
amount created in stage I, while the initial amount of nucleotide in stage II
is whatever was left over from the previous cycle of stage II.
For the model without Taq dynamics,
upon completion of stage II any unextended primed ssDNA will break-up
in dissociation, and thus the initial amount of ssDNA in stage I is
that plus the amount left-over from the previous stage I, plus twice the amount of double
stranded DNA created in stage II.
The initial amount of primer in stage I is also the sum of the
dissociated amount from stage II and the amount leftover from
the previous stage I.  The initial amount of double-stranded DNA
in stage II is assumed to be zero at the start of every cycle.
Written as equations these initial conditions are:

Stage I:
\begin{equation}
s(0)=s(t_{end}, \mbox{previous stage I}) + s'(t_{end},\mbox{previous stage II})+
2 d(t_{end},\mbox{previous stage II});
\end{equation}
\begin{equation}
p(0)=s'(t_{end},\mbox{previous stage II})+
p(t_{end},\mbox{previous stage I});\;\;
s'(0)=0.0;
\label{IIC1}
\end{equation}

Stage II:
\begin{equation}
s'(0)=s'(t_{end},\mbox{ stage I});\;\;\;
n(0)=n(t_{end},\mbox{previous stage II});\;\;d(0)=0.0.
\label{IIC2}
\end{equation}

There are two conserved quantities in the equations for both
stage I and stage II, so the systems can be reduced to one
equation each and solved analytically.  For stage I we use
the conserved quantities: $p(t)-s(t)=K=p(0)-s(0)$ and
$s(t)+s'(t)=K_s=s(0)-s'(0)=s(0)$, (since $s'(0)=0$) so that the resulting
ODE for $p(t)$ is
\[
\frac{dp}{dt}=k_1 (K-p) p,
\]
with the solution
\[
p(t)=\frac{K}{1-\frac{s(0)}{p(0)}\exp(-K k_1 t)}.
\]
and the other variables are found from the conserved quantities:
\[
s(t)=p(t)-K=p(t)-p(0)+s(0),
\]
\[
s'(t)=K_s-s(t)=s(0)-s(t).
\]
It is expedient to rescale the dependent variables in the original
ODEs by the amount of one quantity at the beginning of the reaction,
and since the nucleotides are present in excess we call that $N_0$
and define new variables: $\hat{p}=\frac{p}{N_0},
\hat{s}=\frac{s}{N_0}, \hat{s'}=\frac{s'}{N_0},
\hat{n}=\frac{n}{N_0}, \hat{d}=\frac{d}{N_0}$. We also rescale time
by define a new time $\hat{t_1}=\hat{k_1} t$, where $\hat{k}_1=k_1
N_0$.

The solution is then
\[
\hat{p}(\hat{t}_1)=\frac{\hat{K}}{1-\frac{\hat{s}(0)}{\hat{p}(0)}\exp(-\hat{K} \hat{t}_1 )},
\]
\[
\hat{s}(\hat{t}_1)=\hat{p}(\hat{t}_1)-\hat{K}=\hat{p}(\hat{t}_1)-\hat{p}(0)+\hat{s}(0),
\]
\[
\hat{s}'(\hat{t}_1)=\hat{K}_s-\hat{s}(\hat{t}_1)=\hat{s}(0)-\hat{s}(\hat{t}_1).
\]
where $\hat{K}=\hat{p}(0)-\hat{s}(0)$, $\hat{K}_s=\hat{s}(0)$.

The quantities in stage II can be computed in an identical
manner, and with the choice of conserved quantities $s'(t)+d(t)=K_d=s'(0)+d(0)=s'(0)$,
$n(t)-s'(t)=K_n=n(0)-s'(0)$ the solution is
\[
n(t)=\frac{K_n}{1-\frac{s'(0)}{n(0)}\exp(-K_n k_2 t)},
\]
and the other variables are found from the conserved quantities:
\[
s'(t)=n(t)-K_n=n(t)-n(0)+s'(0),
\]
\[
d(t)=K_d-s'(t)=s'(0)-(n(t)-n(0)+s'(0))=n(0)-n(t).
\]
Rescaling again by the amount of nucleotide at the beginning of the
first cycle, $N_0$, and defining $\hat{t}_2 = \hat{k}_2 t$, (where
$\hat{k}_2=k_2 N_0$) results in
\[
\hat{n}(\hat{t_2})=\frac{\hat{K}_n}{1-\frac{\hat{s}'(0)}{\hat{n}(0)}\exp(-\hat{K}_n \hat{t_2})}
\]
\[
\hat{s}'(\hat{t_2})=\hat{n}(\hat{t_2})-\hat{K}_n=\hat{n}(\hat{t_2})-\hat{n}(0)+\hat{s}'(0)
\]
\[
\hat{d}(\hat{t_2})=\hat{K}_d-\hat{s}'(\hat{t_2})=\hat{n}(0)-\hat{n}(\hat{t_2})
\]
where $\hat{K}_d=\hat{s}'(0)$, $\hat{K}_n=\hat{n}(0)-\hat{s'}(0)$.
The initial conditions can be written in terms of the rescaled variables,
they are identical in form to (\ref{IIC1}) and (\ref{IIC2}), with $\hat{X}$
replacing $X$, for each variable.
From this point forward we rename $\hat{X}= {X}$ to
simplify the notation, while keeping in mind how the rescaling
changes the initial conditions.

A map from one cycle to the next can be constructed from these
solutions and the initial conditions (\ref{IIC1}) and  (\ref{IIC2}).
To distinguish between the concentration of primed single stranded
DNA (ssDNA') in the first and second stage we designate them $s'^{I}$ and
$s'^{II}$.
Let the final time for each stage be fixed at $\tau_I$ and $\tau_{II}$
respectively.  The map for stage I is then:
\[
p_i(\tau_I)=\frac{K_i}{1-\frac{s_i(0)}{p_i(0)}\exp(-K_i \tau_I)},
\]
where $K_i=p_i(0)-s_i(0)$, and
\[
s_i(\tau_I)=p_i(\tau_I)-K=p_i(\tau_I)-p_i(0)+s_i(0),
\]
\[
s'^{I}_i(\tau_I)=K_s-s_i(\tau_I)=s_i(0)-s_i(\tau_I),
\]
and for stage II:
\[
n_i(\tau_{II})=\frac{K_n}{1-\frac{s'^{II}_i(0)}{n_i(0)}\exp(-K_n \tau_{II})},
\]
where $K_n=n_i(0)-s'^{II}_i(0)$,
and
\[
s'_i(\tau_{II})=n_i(\tau_{II})-K_n=n_i(\tau_{II})-n_i(0),
\]
\[
d_i(\tau_{II})=K_d-s'^{II}_i(\tau_{II})=n_i(0)-n_i(\tau_{II}).
\]

The initial conditions for the $i$th cycle are:
\[
s_i(0)=s_{i-1}(\tau_I)+s'^{II}_{i-1}(\tau_{II})+ 2 d_{i-1}(\tau_{II}),
\]
\[
p_i(0)=p_{i-1}(\tau_I)+s'^{II}(\tau_{II}),
\]
\[
s'^{I}(0)=0.0,
\]
\[
s'^{II}_i(0)=s'^{I}_i(\tau_I),
\]
\[
n_i(0)=n_{i-1}(\tau_{II}),
\]
\[
d_i(0)=0.0.
\]

While a closed form version of the map can be written down, it is
not particularly illuminating, simulations must be performed to
uncover the behavior of the solutions.  To do this parameters must
be estimated or fit from the data, these are $\tau_I, \tau_{II}$,
and the initial concentrations of primer and ssDNA, $p_1(0),
s_1(0)$, relative to the initial concentration of nucleotides,
$N_0$. The model and data must both saturate at the same level, and
a scaling quantity for the data values was fit for each curve, which
we called $KK$.  The parameter estimation was done as in the
previous section, with the same set of data.   The percentage error
in each fit was computed by dividing the final value of the
objective function by the two-norm of the model time series vector.

\begin{figure}
\centerline{\epsfig{file=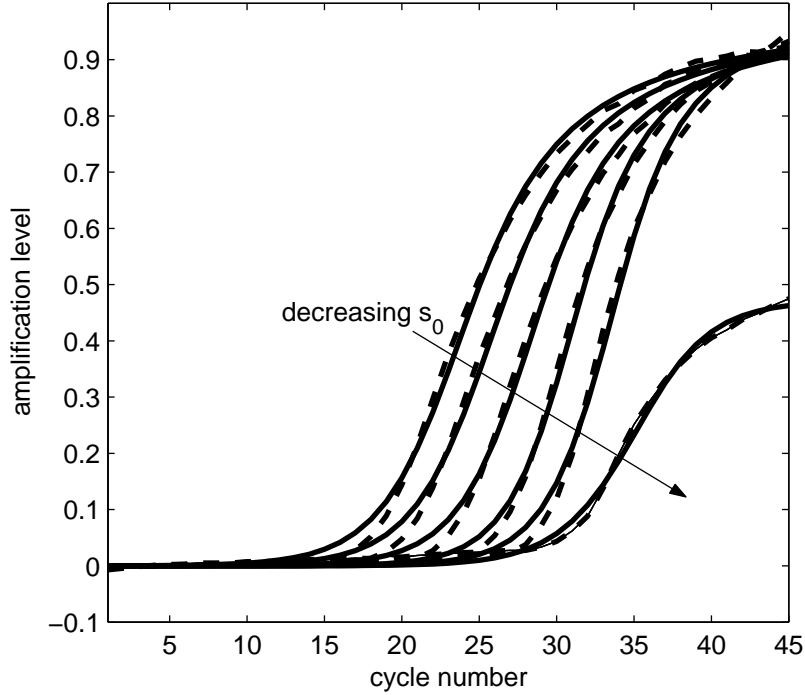}}
\caption{Fitting dilution series data to the two stage model without
Taq dynamics.  For parameter values see Table IVa.}
\label{intmodelfit}
\end{figure}

 For the dilution series all these
parameters were fit,  see figure \ref{intmodelfit} and Table IVa.
The $KK$ value found is consistent with the variation of the
saturation value for each run, and for each run $\tau_1$ is greater
than $\tau_2$, indicating that annealing is slower than extension.
The initial quantity $s_1(0)$ decreases roughly by a factor of 10
for each run, also consistent with the dilution series. The initial
amount of primer in each run is close to 1.0, indicating the amount
of primer must be close to the amount of nucleotide in a run to
match the data, which is not consistent with the information we have
about the experiment, which is that the nucleotides (measured in
blocks of the template sequence to be amplified) to primer ratio is
about 4-to-1. We return to this issue when we examine the model
including Taq dynamics.  The 6th run has a much larger error than
the others, indicating that other factors come into play in the
dynamics of very small copy number runs, such as competition from
other species (e.g. primer-dimers).

We then fit the model to the runs with unknown initial
concentrations, of which there are two, each with five replicates.
For these 10 runs $\tau_I, \tau_{II}, p_1(0)$ were fixed at values
from the dilution run that comes out of background at nearly the
same time (for the sample 1 this was run 2, for sample 2, run 4).
The results from this exercise are presented in Table IVb.

\begin{center}
Table IVa: Parameter Estimation of the Two Stage Model without Taq Dynamics-Dilution Series

\begin{tabular}{|l|c|c|c|c|c|c|}        \hline
    &KK & $\tau_I$  & $\tau_{II}$   & $s_1(0)$  & $p_1(0)$  & \% error \\ \hline
run 1  & 8.989  & 1.525 & 0.829 &   1.84e-04 &  0.9935 &   10.1 \\  \hline
run 2 & 9.169 &   1.551 &  0.853 &  7.23e-05 &  0.9933 & 9.8 \\ \hline

run 3  & 8.768 & 1.597 & 0.916  & 1.67e-05 & 0.9930 & 9.4 \\ \hline

run 4 & 7.390  & 1.466 & 1.216 & 1.70e-06 & 0.9978 & 8.0 \\ \hline

run  5  & 4.902  & 1.454 &   1.598 &  1.68e-07 & 0.9636 & 10.4 \\ \hline

run 6  & 1.645  & 4.787 &  0.939 & 8.00e-8 & 0.991 & 19.38 \\ \hline
\end{tabular}
\end{center}

\begin{center}
Table IVb: Parameter Estimation of Two Stage Model w/out Taq Dyn.-Unknown Samples
\begin{tabular}{|l|c|c|c|}      \hline
    & $s_1(0)$ & KK & \% error\\ \hline
sample 1a &   8.019e-05  & 10.35 &     14.7 \\ \hline
sample 1b & 7.87e-05 & 10.482 & 13.0 \\ \hline
sample 1c & 7.67e-05 &  10.518 &    11.2 \\  \hline
sample 1d &  7.57e-05  & 10.713 &    10.9 \\ \hline
sample 1e & 7.42e-05 &   10.722 &    10.7 \\ \hline
sample 2a & 1.51e-06 &  7.158  &  9.4 \\ \hline
sample 2b & 1.40e-06 &  7.490 &  8.6 \\ \hline
sample 2c & 1.40e-06  & 7.518 &  8.6 \\ \hline
sample 2d & 1.39e-06 &   7.688  &   8.5 \\ \hline
sample 2e & 1.50e-06 &   7.430 &   8.8 \\ \hline
\end{tabular}
\end{center}

\subsection{Two Stage Model with Taq Dynamics}

For this model the equations for stage I remain as (\ref{a1})-(\ref{a3}),
while the stage II equations now include both Taq and complex concentrations
and are given by  (\ref{ee1})-(\ref{ee3}).   As in the simpler model
the initial amount of primed ssDNA in stage II is equal to the
amount created in stage I, while the initial amount of nucleotide in stage II
is whatever was left over from the previous cycle of stage II.
Upon completion of stage II any unextended complex will break-up
during dissociation, as will any primed ssDNA.
 Thus the initial amount of complex in stage II will be zero, and
the amount of Taq enzyme will be the original amount from the
beginning of the PCR reaction ($Q$).   Stage I starts with no primed
ssDNA, it is assumed to dissociate during the melt phase. The primer
initial condition is the amount of unused primer from the previous
cycle, plus the amount created during the dissociation of the
complex during the melt phase. The initial amount of ssDNA will be
that left from the previous annealing phase plus an amount equal to
amount of complex left in stage II that dissociates, plus the ssDNA
that results from the dissociation of the dsDNA created in the
previous stage II, which is double the amount of dsDNA. In terms of
equations these initial conditions can be stated:

\begin{equation}
s'(0)=s'(t_{end},\mbox{ stage I});\;\;q(0)= Q ;\;\;c(0)=0.0;\;\;
n(0)=n(t_{end},\mbox{previous stage II});\;\;d(0)=0.0.
\label{IC2}
\end{equation}

and for stage I:
\begin{equation}
s(0)=s(t_{end}, \mbox{previous stage I}) + c(t_{end},\mbox{previous stage II}) +
2 d(t_{end},\mbox{previous stage II});
\label{IC1}
\end{equation}
\[
p(0)=c(t_{end},\mbox{previous stage II}) + s'(t_{end},\mbox{previous stage II})+
p(t_{end},\mbox{previous stage I});
\]
\[
s'(0)=0.0.
\]

A map for the reaction is created by integrating the ODEs in each
stage and using the initial condition rules to link one stage
to the other. However, insight can be gained by analyzing
the dynamics of each stage separately  and forming some special limiting
cases for this map.

\subsubsection{Dynamics of Stage I}
The equations for stage I are again completely integrable, because of two
conserved quantities,  $s+s'=K_s=s'(0)+s(0)=s(0)$, and $p-s=K=p(0)-s(0)$.
Simplifying the equation for $p$ using these quantities yields
\begin{equation}
\frac{dp}{dt}=k_1 (K-p) p.
\label{stage1ode}
\end{equation}
Rescaling time, $\tilde{t}=p(0) k_1 t, \frac{dp}{d\tilde{t}}=\dot{p}$,
and the molecular concentration of all three quantities by $p(0)$, i.e.
$\hat{p}=\frac{p}{p(0)}, \hat{s}=\frac{s}{p(0)}, \hat{s'}=\frac{s'}{p(0)}$,
and $\hat{K}=\frac{p(0)-s(0)}{p(0)}=1-\hat{s}(0)$,
results in
\[
\dot{\hat{p}}=(\hat{K}-\hat{p}) \hat{p},
\]
which has  the solution
\begin{equation}
\hat{p}(\tilde{t})=\frac{\hat{K}}{1-\hat{s}(0)e^{-\hat{K} \tilde{t}}},
\label{stage1soln}
\end{equation}
so the remaining quantities can be computed
\[
\hat{s}(\tilde{t})=\hat{p}(\tilde{t})-\hat{K}=\hat{p}(\tilde{t})+\hat{s}(0)-1,
\]
and
\[
\hat{s}'(\tilde{t})=\hat{K}_s-\hat{s}(\tilde{t})=1-\hat{p}(\tilde{t}).
\]
Note that in the limit as $\tilde{t}\rightarrow \infty$,
$\hat{p}(\tilde{t}) \rightarrow \hat{K}$ if $\hat{K} > 0$, e.g. $1 >
\hat{s}(0)$, more initial primer than ssDNA, and $\hat{p}(\tilde{t})
\rightarrow 0$ if $\hat{K} < 0$, i.e., there is more ssDNA to begin
with than primer.   There is a transcritical bifurcation at
$\hat{K}=0$, $\hat{s}(0)=1$,  where the two fixed points for the
system (\ref{stage1ode}), $\bar{p}=K$ and $\bar{p}=0$, exchange
stability.

\subsubsection{Dynamics of Stage II}
The stage II ODEs can be simplified by rescaling time to remove one rate  constant.
The dimensionless time chosen, $\tau$,  is $k_{-2} t$ and the new system is
\begin{equation}
 \dot{s'}=-\gamma s' q + c
\label{E1}
\end{equation}
\begin{equation}
\dot{q}=-\gamma s' q+  c+ \beta c n
\label{E2}
\end{equation}
\begin{equation}
\dot{c}=\gamma s' q- c- \beta c n
\label{E3}
\end{equation}
\begin{equation}
\dot{n}=-\beta c n
\label{E4}
\end{equation}
\begin{equation}
\dot{d}=\beta c n
\label{E5}
\end{equation}
where $\frac{df}{d\tau} = \dot{f}, \beta=\frac{k_3}{k_{-2}},
\gamma=\frac{k_2}{k_{-2}}$.

The initial conditions are (as stated previously):
\[
s'(0)=s'(t_{end},\mbox{stage I});\;\;q(0)=Q;\;\;c(0)=0.0;\;\;n(0)=n(t_{end},
\mbox{previous stage II}),\;\; d(0)=0.0.
\]
The molecular quantities in the above equations can themselves be
scaled, we choose here to scale by the initial amount of nucleotides
available at the start of the cycle, $N_0=n(0)$.  The rescaled
equations have the same form, with rescaled parameters
$\gamma=\gamma/N_0$ and $\beta=\beta/N_0$ (meaning we adopt the
notation $\hat{X}=\frac{X}{N_0}$ for each quantity, and then discard
the hat for simplicity). The corresponding initial conditions are
\[
s'(0)=\frac{s'(t_{end},\mbox{stage I})}{N_0};\;\;q(0)=\frac{Q}{N_0};\;\;c(0)=0.0;\;\;n(0)=1,\;\; d(0)=0.0.
\]

The equations (\ref{E1})-(\ref{E5}), have three conserved quantities (e.g. $n+d, q+c, s+c+d$),
so the dynamics can be reduced from a five to a
two-dimensional system.  Since initially there will be no $c$ or $d$, the
these conserved quantities can be written  $s'+c+d=s'(0), \;\;q+c=q(0),\;\; n+d=1$.
The two dimensional system that results from incorporating the conserved
quantities is:
\[
\dot{q}=-\gamma(s'(0)-q(0)-1+q+n)q+q(0)-q+\beta (q(0)-q)n
\]
\[
\dot{n}=-\beta(q(0)-q)n.
\]
From this we can more readily determine fixed points and analyze their
stability.   Two of the three fixed points for this system are physically
relevant, and are given by
\[
\mbox{f.p. 1} = (\bar{s'}=0,\;\; \bar{q}=q(0),\;\; \bar{c}=0,\;\;
\bar{n}=1-s'(0), \;\; \bar{d}=s'(0)),
\]
and
\[
\mbox{f.p. 2}= (\bar{q}=\frac{1}{2\gamma}(\gamma(q(0)+1-s'(0))-1+
\sqrt{(-\gamma(q(0)+1-s'(0))+1)^2+4\gamma q(0)})),
\]
\[
\bar{s'}=s'(0)-q(0)-1+\bar{q},\;\; \bar{c}=q(0)-\bar{q},\;\;
\bar{n}=0,\;\; \bar{d}=1).
\]
The third fixed point has the $q$-coordinate:
$$\bar{q}=\frac{1}{2\gamma}(\gamma(q(0)+1-s'(0))-1- \sqrt{(-\gamma(q(0)+1-s'(0))+1)^2+4\gamma q(0)}))$$
which can be shown to be always negative, and so physically
irrelevant.  For a proof of this fact, and the complete stability
analysis, see appendix I. Here we state the main result only: There
is a transcritical bifurcation when $s'(0)=1=n(0)$, here f.p. 1 and
f.p. 2 are identical and exchange stability.  For $n(0) > s'(0)$
f.p. 1 is attracting and has non-negative coordinate values. For
$n(0) >  s'(0)$  f.p. 2 is attracting with non-negative coordinate
values.   The reaction will typically begin with more nucleotides
than primed ssDNA, so the long-time behavior is represented by f.p.
1, which has final dsDNA value equal to the amount of primed ssDNA
(all primed ssDNA is extended). If the initial amount of primed
ssDNA exceeds the amount nucleotide at the start of that cycle, the
second fixed point becomes attracting and non-negative, and in this
case the long term dsDNA amount will be equal to the initial amount
of nucleotide.  Hence the limiting value of dsDNA switches from the
initial amount of primed ssDNA to the initial amount of nucleotide
as primer becomes limiting, as would be anticipated.

\subsection{Limiting Cases}
Assuming both stage I and stage II ODEs reach steady state, a map
can be constructed by linking the attracting fixed points through
the initial conditions. Since the rescaling of each stage used in
the previous subsection is different, (and changes with each cycle) it is best to return to the
unscaled quantities to create the map.  The fixed points for stage II
in terms of the unscaled quantities are the same, with the 1.0's replaced
by $n(0)$'s.  The bifurcation occurs when $s'(0)=n(0)$, as was outlined
previously.

The attracting fixed point for stage I depends on the initial values
of primer and ssDNA, we will write it
$(\bar{p},\;\bar{s},\;\bar{s}^{II}= p(0)-s(0),\;0,\;s(0))$ if
$p(0)>s(0)$ (there is enough primer to prime all the ssDNA). If
$p(0)<s(0)$ it is $(\bar{p},\;\bar{s},\;\bar{s}'^{II}=
0,\;p(0)-s(0),\;p(0))$: the limiting amount is primer.   The
attracting fixed point for stage II depends on the relative size of
$n(0)$ and $s'(0)$, the latter being equal to $\bar{s'}$ from the
previous stage I.   The initial amount of nucleotide will be
determined by how much remains from the previous cycle.   In a
similar manner the rest of the initial conditions are determined by
the fixed points from the previous cycle, this dependance is
detailed below.

For stage I the initial amount of primer is the sum of what is left over
from the previous stage I (cycle $i-1$), and the primer released from
the dissociation of the complex and the unextended primed
ssDNA from the previous stage II:
\[
p_i(0)=\bar{p}_{i-1}+\bar{c}_{i-1}+\bar{s}'^{II}_{i-1}.
\]
The initial amount of ssDNA will be twice the amount of dsDNA
created in the previous cycle, plus the ssDNA released from
the dissociation of the complex and the unextended primed
ssDNA from the previous stage II, plus the amount of ssDNA
not primed in the previous stage I:
\[
s_i(0)=2 \bar{d}_{i-1}+\bar{c}_{i-1}+\bar{s}'^{II}_{i-1}+\bar{s}_{i-1}.
\]
At the start of stage I there is no primed ssDNA:
$s'^{I}_i(0)=0.$

In stage II the initial amount of primed ssDNA is equal to the
amount coming out of stage I, $\bar{s}'^I_i$, the initial amount of
Taq is a constant, $Q$, the complex has been completely dissociated
during the melt phase, along with any dsDNA.  The amount of
nucleotide is equal to the amount left over from the previous stage
II, so the ICs are:
\[
s'^{II}(0) = \bar{s}'^I_i;\;\; q_i(0)=Q;\;\; c_i(0)=0.0;\;\;
n_i(0)=\bar{n}_{i-1};\;\; d_i(0)=0.0.
\]

There will be four distinct cases of the map depending on
the relative size of $n(0), s'^{II}(0)$ and $p(0), s(0)$. Here we limit
the analysis to a physically realistic scenario:  initially both
primer and nucleotide dominate, but initial primer amount is less
than initial nucleotide amount.  In the course of creating new amplicon
both primer and nucleotide amounts decrease, and since they are used
in the same proportion the primers will be exhausted before the nucleotides.
This will cause a shift to another case of the map, and the remaining
two iterations will complete the process, since no more primer will
be available to make the extension possible.  The equations for this
scenario are presented next.

First assume $p_i(0) > s_i(0)$ and $n_i(0) > s-s'^{II}(0)$, so the
stage I fixed point is (for the $i$th cycle)
\[
\bar{p}_i=p(0)_i-s(0)_i;\;\;\bar{s}_i=0;\;\;\bar{s}_i'^{I}=s_i(0).
\]
Then assume $n_i(0)>s_i'^{II}(0)$ so that the fixed point for stage
II is
\[
\bar{s}'^{II}_i=\bar{c}_i=0,\;\;\bar{q}_i=q(0).
\]
\[
\bar{n}_i=n_i(0)-s_i'^{II}(0),
\]
\[
\bar{d}_i=s_i'^{II}(0).
\]
Now, the initial conditions for cycle $i$ are determined by
the fixed points from cycle $i-1$ in this manner:
\[
p_i(0)=
\bar{p}_{i-1}+\bar{c}_{i-1}+\bar{s}'^{II}_{i-1}=\bar{p}_{i-1}
\]
\[
s_i(0)=2\bar{d}_{i-1}+\bar{c}_{i-1}+\bar{s}'^{II}_{i-1}+\bar{s}_{i-1}=2
d_{i-1}.
\]
Substituting these into the fixed point for stage I yields:
\[
\bar{p}_i=\bar{p}_{i-1}-2 \bar{d}_{i-1};\;\;
\bar{s}_i=0;\;\;
\bar{s}'^{I}=2 \bar{d}_{i-1}.
\]
The initial conditions for stage II in terms of fixed points
for the previous cycle are
\[
n_i(0)=\bar{n}_{i-1};\;\; s_i'^{II}(0)=\bar{s}_i'^{I}=2
\bar{d}_{i-1}.
\]
And so the stage II fixed point is then
\[
\bar{s}'^{II}_i=\bar{c}_i=0,\;\;\bar{q}_i=Q
\]
\[
\bar{n}_i=\bar{n}_{i-1}-2 \bar{d}_{i-1}
\]
\[
\bar{d}_i=2 \bar{d}_{i-1}
\]

This is simple doubling of the double-stranded DNA, and will proceed
until the amount of primer ($p(0)$) at the beginning of stage I is less
than the amount of single-stranded DNA ($s(0)$).  At this cycle (call it $N$) the
attracting fixed point switches in stage I and a new map is created that is
valid for exactly one cycle.   The new stage I fixed point is:
\[
\bar{p}_N=0
\]
\[
\bar{s}_N=s_N(0)-p_N(0)=2 \bar{d}_{N-1}-\bar{p}_{N-1}
\]
\[
\bar{s}'^{I}_N=p_N(0)=\bar{p}_{N-1}
\]
In stage II the amount of nucleotide is depleted by an amount equal
to the amount of $\bar{s}'^{I}$ created, which equals $\bar{p}_{N-1}$.
The amount of double stranded DNA will be equal to that amount as well,
so the map for stage II is:
\[
\bar{s}'^{II}_N=\bar{c}_N=0,\;\;\bar{q}_N=q(0)
\]
\[
\bar{n}_N=\bar{n}_{N-1} - \bar{p}_{N-1}
\]
\[
\bar{d}_N= \bar{p}_{N-1}
\]
For the next cycle, $N+1$, there is no primer left so the duplication ends.
The fixed point values in this cycle are:
\[
\bar{p}_{N+1}=0
\]
\[
\bar{s}_{N+1}=2 \bar{d}_{N}+\bar{s}_N=2 \bar{p}_{N-1}+2\bar{d}_{N-1}-\bar{p}_{N-1}=
\bar{p}_{N-1}+2 \bar{d}_{N-1}=s_{\mbox{final}}
\]
\[
\bar{s}'^{I}_{N+1}=0.
\]
For stage II:
\[
\bar{n}_{N+1}=\bar{n}_N-\bar{p}_N=\bar{n}_{N-1}-\bar{p}_{N-1}=n_{\mbox{final}}
\]
\[
\bar{d}_{N+1}=0
\]
It is straight-forward to show that this is a fixed point for the
map from one cycle to the next, once this stage is reached the final
value of extended DNA is fixed.   It lives in the reaction as single
stranded DNA until the mixture is allowed to ``finish off" and the
strands reanneal (finishing off more technically refers to the stage
in which the extension is allowed to run to completion, and all the
primed single-stranded DNA molecules have been turned into double
stranded DNA, which happens in cycle $N$).

This map, created for the limiting case of infinite time for
each stage in each cycle, converges on the model of simple doubling
until the reaction limiting species  (either $p$ or $n$)
is exhausted.   Then in one cycle the reaction finishes off and
reaches  a fixed point.  Clearly this does not capture the
sigmoidal growth curve or the variation away from it.
 The next step is to allow the extension in stage II to
reach the asymptotic fixed point determined by initial conditions
for each cycle, but to use the exact solution of the stage I
equations, with the run time left as a parameter, to determine the
values at the end of stage I. We construct this map next.

The exact solution for the stage I variables is as follows:
\begin{equation}
p_i(T_I)=\frac{K}{1+\frac{s_i(0)}{p_i(0)}e^{-K T_I}}= f_{T_I}(p_i(0),s_i(0)),
\end{equation}
where $K=p_i(0)-s_i(0)$.  The unprimed and primed ssDNA depend on $p$
through the conserved quantities:
\begin{equation}
s_i(T_I)=p_i(T_I)+s_i(0)-p_i(0)
\end{equation}
\begin{equation}
s_i'^I(T_I)=p_i(0)-p_i(T_I).
\end{equation}
The stage II fixed point for $n(0)>s'^{II}(0)>0$ will  be:
\[
\bar{s}'^{II}_i=\bar{c}_i=0,\;\;\bar{q}_i=Q
\]
\[
\bar{n}_i=n_i(0)-s_i'^{II}(0)=n_i(0)-s_i'^{I}(T_I)
\]
\[
\bar{d}_i=s'^{II}(0)=s_i'^{I}(T_I)=p_i(0)-p_i(T_I).
\]
The initial conditions for the next stage I are then
\begin{equation}
p_{i+1}(0)=p_i(T_I)+\bar{c}_i+\bar{s}'^{II}_i=p_i(T_I),
\end{equation}
\begin{equation}
s_{i+1}(0)=2 \bar{d}_i+\bar{c}_i+\bar{s}'^{II}_i+s_i(T_I)
=2(p_i(0)-p_i(T_I))+p_i(T_I)+s_i(0)-p_i(0)\\
=p_i(0)-p_i(T_I)+s_i(0),
\end{equation}
and
\begin{equation}
s'_{i+1}(0)=0.
\end{equation}
Note that at this point the stage I initial conditions, in which the
map is cast, depend only on the previous stage I values, the stage I
map runs independently of stage II.
The stage II fixed point can be determined directly from the stage I
variables, and the only one of interest is the
amount of nucleotide, for when that is exhausted the reaction will
stop.  The equation for nucleotide, $n$, is:
\begin{equation}
n_{i+1}(0)=\bar{n}_i=n_i(0)-s_i'^{I}(T_I)=n_i(0)-(p_i(0)-p_i(T_I)).
\end{equation}

There are two possibilities for the completion of the reaction:
either primer runs out or the nucleotides.  In the case that primer
runs out first, we look at the limit as $p_i(0) \rightarrow 0$, so
that $p_i(T_i) \rightarrow 0$ and $s_i(T_I) \rightarrow s_i(0)$ and
$s'_i(T_I) \rightarrow 0$.   In stage II no complex will be formed,
or double-stranded DNA created, as there is no primed ssDNA
available at the start of the reaction: $s'^{II}_i(0)=s'_i(T_I)=0$.
The fixed point for the nucleotide is thus the value of the
nucleotide at the beginning of the cycle,
$\bar{n}_i=n_{i}(0)-s'^{I}_{i}(T_I)=n_{i}(0)$, and the map for
$n_{i}$ is at a fixed point, $n_{i+1}(0)=n_{i}(0)$.

If the resource is the limiting factor, rather than primer, in the
second to last cycle ($(r-1)$-cycle) we have $n_{r-1}(0) <
s'^{II}_{r-1}(0)$. This sends stage II variables to f.p. 2. In the
final stage I the initial conditions are then
\[
p_r(0)=p_{r-1}(T_I)+\bar{c}_{r-1}+\bar{s}'^{II}_{r-1}
\]
\[
s_r(0)=2
\bar{d}_{r-1}+\bar{c}_{r-1}+\bar{s}'^{II}_{r-1}+s_{r-1}(T_I),
\]
with f.p. 2 values.
At the end of stage I the function values are:
\[
p_r(T_I)=f_{T_I};\;\;s_r(T_I)=f_{T_I}+s_r(0)-p_r(0);\;\;s'^{I}_r(T_I)=p_r(0)-f_{T_I}.
\]
The initial condition for $s'^{II}_{r}(0)=s'^{I}_r(T_I)=p_r(0)-f_{T_I}$,
but only complex can be formed in the final stage II, since
the resources have been exhausted, $n_r(0)=0$.  During the
dissociation phase the complex breaks up, and the initial quantities
for stage I are $p_{r+1}(0)=p_{r}(0);\;\; s_{r+1}(0)=s_r(0)$.
The reaction has reached a fixed point with ($\bar{p}=p_{r}(0),\;\;\bar{s}=s_r(0),
\;\;\bar{n}=0$).

\begin{figure}
\centerline{\epsfig{file=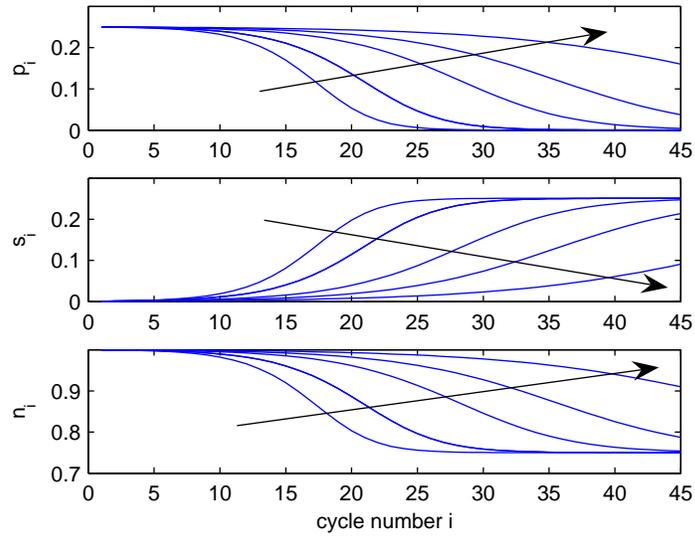}}
\caption{Integration of map with varying stage I
integration time.  Arrows point in the direction of decreasing stage I
integration time.}\label{intermap}
\end{figure}

\begin{figure}
\centerline{\epsfig{file=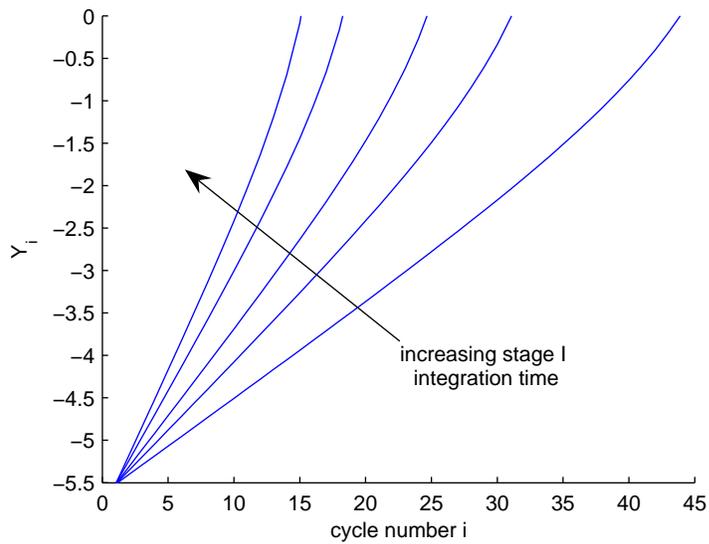}}
\caption{Logarithmic regression variable $Y_n$ from the two-stage map with
varying stage I integration time.}\label{intmaplogreg}
\end{figure}

To illustrate the dynamics of this map we plot $n_i, s_i$ and $p_i$
in figure \ref{intermap}. Six runs were performed with varying stage
I integration time: ($t_I=0.5,0.75,1.0,1.5,2.0$). The initial
conditions are $n_0= 1.0; p_0=0.25; s_0=0.001$, i.e., the case in
which primer is limiting. To further analyze the effect of varying
integration time we plot the logarithmic regression of $s$, the
quantity $Y_i = \ln( |\frac{s_i}{p_0-s_i}|)$, in figure
\ref{intmaplogreg}. Here we see that in the limit of shorter
integration times the growth is more nearly logistic,  and for
longer integration times it deviates from logistic by being concave
up, rather than concave down, which is what is seen in the
amplification data (figure \ref{logreg}).   Including variation in
stage I integration time is clearly not enough to capture the
correct non-sigmoidal behavior of the growth curves, some other part
of the reaction dynamics must explain the accentuated slowing of
growth during the latter half of the reaction. This leads us to
integrating and fitting parameters on the full model, eq.s
(\ref{a1})-(\ref{a3}), and (\ref{e1})-(\ref{e5}).

\subsection{Parameterizing the Full Two Stage Model with PCR Data}

We now investigate the parameterizations of the model with arbitrary
time in stage I and in stage II. The reactions in the annealing
phase (stage I) are the same those presented in equations
(\ref{stage1ode}),(\ref{stage1soln}), and the linking initial
conditions are  (\ref{IC1}). The stage II ODEs are given in
equations (\ref{E1})-(\ref{E5}) with rescaled parameters and initial
conditions:
\[
s'(0)=\frac{s'(t_{end},\mbox{stage I})}{N_0};\;\;q(0)=\frac{Q}{N_0};\;\;c(0)=0.0;\;\;n(0)=1.0,\;\; d(0)=0.0.
\]

\begin{figure}
\centerline{\epsfig{file=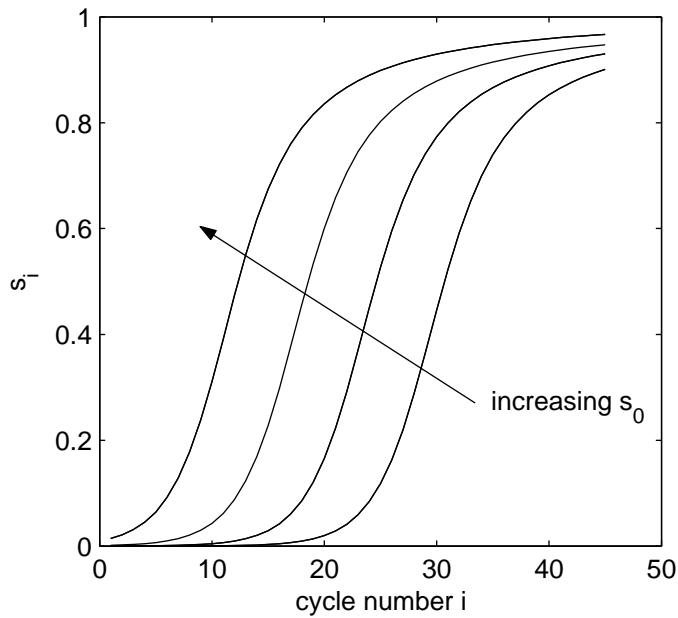}}
\caption{Integration of the coupled ODE model, with varying initial template amount,
$s_0(0)$.}
\label{odemodel}
\end{figure}

Integrating the complete model occurs in phases, first computing the
solution of stage I, eq. (\ref{stage1soln}), then the value of the
stage I variables at $T_I$ are used as initial conditions eq.
(\ref{IC2}) for the stage II integration, eqs.
(\ref{e1})-(\ref{e5}). The following dissociation phase breaks up
existing complex into ssDNA, primer and Taq, and dsDNA into twice as
many ssDNA strands, and these are used as initial conditions for the
next stage I (see eq.(\ref{IC1})). Examples of runs with varying
amounts of initial template are shown in figure \ref{odemodel}.

The parameter estimation was done using the same quantitative PCR
data set in the previous sections, again using the Matlab function
{\it fminsearch} to minimize the mean square error between the
amplification data and the simulated time series.   In performing
the parameterizations we used the value determined for the initial
$s$ level from fitting the two stage model without Taq dynamics. We
fit the normalization constant for the data, $KK$, the two reaction
coefficients, $\beta$ and $\gamma$, and the reaction times $t_I$ and
$t_{II}$. That leaves the initial amount of primer, $p_1(0)$, and of
taq, $Q$, relative to the initial amount of nucleotides. From the
results of many parameterization runs we determined that the best
fit was obtained when $p_1(0)=1.0$, which is not what is indicated
by IT protocol, where a standard reaction set-up has 0.5 micromole
of each primer and 0.8 millimole of dNTPs, the base pairs (BP) used
in extension.   Given an amplicon of 200 BPs this means about 4
micromole of completed segments, or 2 micromole of each
complementary segment.  The ratio of primers to nucleotides is about
four to one, then, so we should set the initial primer amount to
0.25.   This never achieved the same goodness-of-fit that the runs
with  higher initial amounts of primer did.   On the other hand, the
parameterization was relatively insensitive to the initial amount of
Taq polymerase, which we set at 1.0.   See Tables Va,b for the
parameterization results.

\begin{figure}
\centerline{\epsfig{file=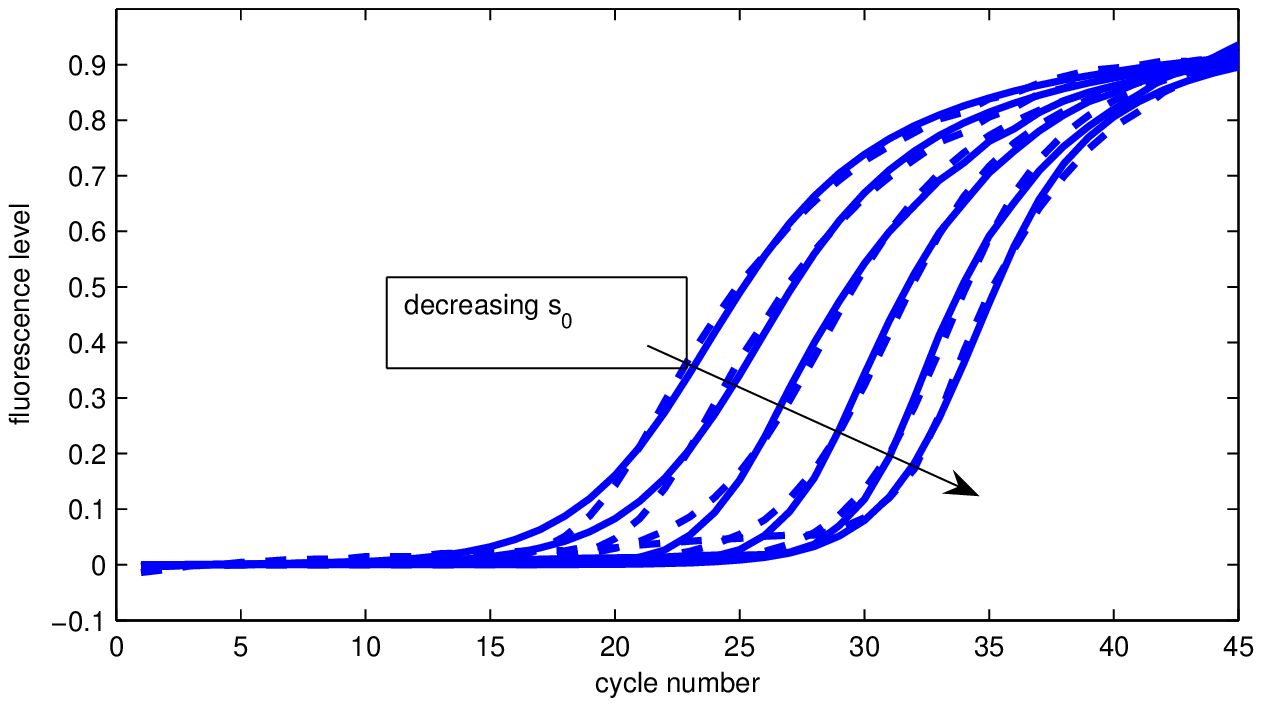}}
\centerline{\epsfig{file=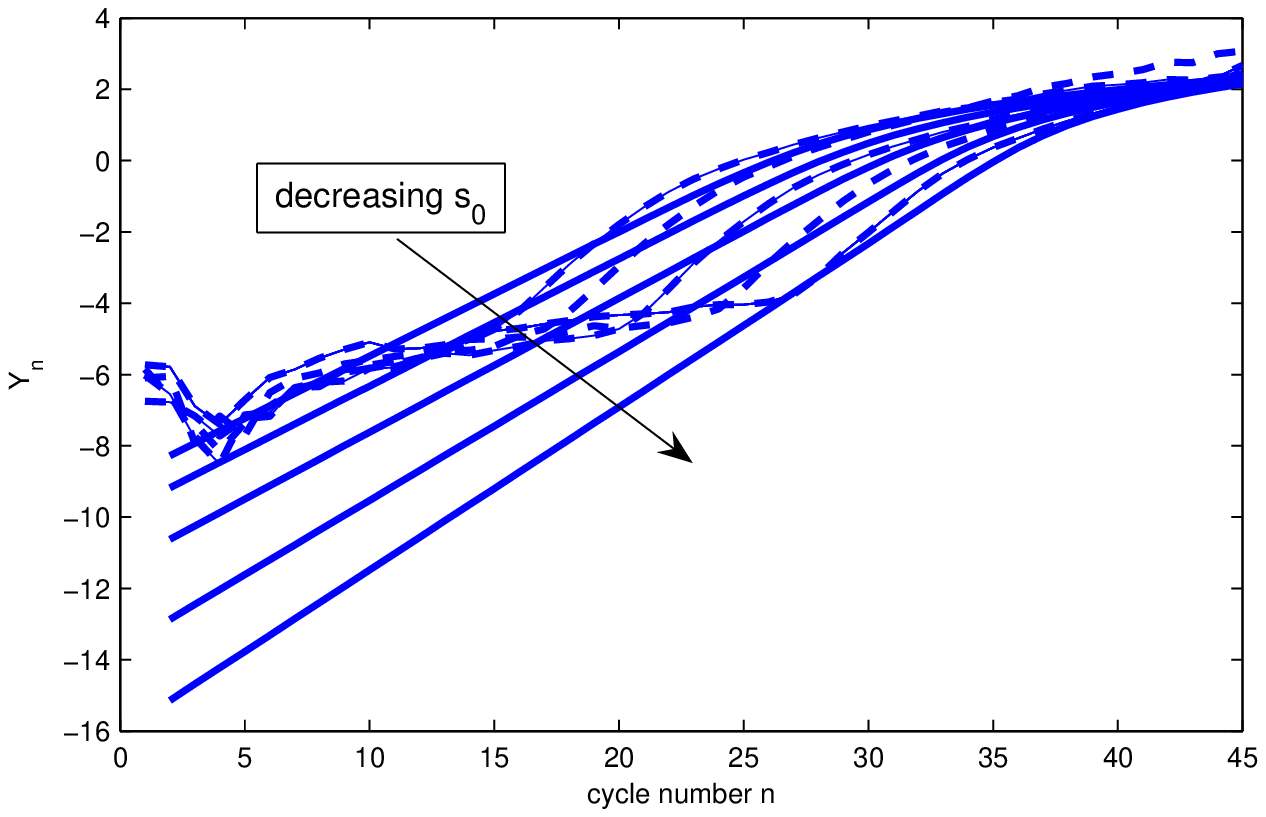}}

\caption{Comparison of two stage model with Taq dynamics to data. a)
amplification curves, b) logarithmic regression curves.  The data is
represented with dashed lines.  See the text for parameter values.}

\label{fittedodemodel}
\end{figure}

 In figure
\ref{fittedodemodel} a) we show a comparison of data to model with
parameters found by the algorithm, for the dilution series.   The
logarithmic regression variable, $Y_i=log(\frac{y_i}{KK-y_i})$ was
plotted for model and the same data in figure \ref{fittedodemodel}
b).

\begin{center}
Table Va: Parameter Estimation: Two Stage Model with Taq Dynamics-Dilution Series

\begin{tabular}{|l|c|c|c|c|c|c|c|}      \hline
    &KK & $s_1(0)$ & $\beta$ & $\gamma$ & $t_I$ & $t_{II}$  & \% error \\ \hline

run 1 & 9.126 &     1.8e-04 &   4.0267 &  0.8650 & 2.2137 &  1.1521  & 2.57 \\ \hline
run 2  & 9.3252 &   7.23e-05      &  5.4532   & 1.0404   & 2.2440 &   0.9242 &  2.62   \\ \hline
run 3   &      8.9122   &      1.67e-05  &  4.8744   &   0.9028  &  2.2876  &  1.1484  &     2.66  \\ \hline
run 4  & 7.5077   &   1.70e-06    & 4.6310  &  0.9777   & 2.5459   & 1.2457   &   2.43  \\ \hline
run 5      &  5.0893   &       1.68e-07  &   4.3792   &  0.9423   &  2.0108  &  1.6888  &   3.05  \\ \hline
run 6      &  1.6993  &        8.00e-08   &  5.9758  &  0.6248  &  1.9002  &  2.3773    & 5.52 \\ \hline

\end{tabular}
\end{center}

We then performed the parameter estimation with data from the
replicates with unknown initial concentrations, the results are
presented in Table  Vb.

\begin{center}
Table Vb: Parameter Estimation: Two Stage Model with Taq Dynamics-Unknown Samples
\begin{tabular}{|l|c|c|c|c|c|c|c|}      \hline
    &KK & $s_1(0)$  & $\beta$   & $\gamma$  & $t_I$ & $t_{II}$  &  \% error \\ \hline

sample 1a & 10.13 &  8.02e-05  & 0.8133  &  1.5316  &  2.5651  &  1.8070  &  3.42\\ \hline
sample 1b & 10.49 & 7.87e-05 & 1.7851  &  0.7927  &  3.3509 &   1.6211   & 3.32 \\ \hline
sample 1c & 10.42 & 7.67e-05 &3.1262  &  2.2279  &  4.3212  &  0.6089    &  2.86  \\  \hline
sample 1d & 10.73  &   7.57e-05  & 4.5382  &  1.3121  &  4.2464  &  0.7309 &  2.86  \\ \hline
sample 1e & 10.82 & 7.42e-05 &  5.4529   & 1.3825  &  3.1713  &  0.6814 &  2.90 \\ \hline

sample 2a & 7.18 & 1.51e-06 & 7.6983  &  1.2537  &  4.7358   & 0.7902    & 2.76\\ \hline
sample 2b & 7.54 & 1.40e-06 & 4.6407  &  0.7659 &   4.2093  &  1.3862      & 2.60\\ \hline
sample 2c & 7.60 & 1.40e-06 & 5.4337  &  0.9113  &  3.6034  &  1.1658     & 2.67\\ \hline
sample 2d & 7.78 & 1.39e-06 &  6.7849  &  0.8755  &  3.4071 &   1.1475      & 2.64\\ \hline
sample 2e & 7.50 & 1.50e-06 &  4.7158  &  0.7127  &  3.6256  &  1.4914     & 2.73\\ \hline

\end{tabular}
\end{center}

Comparing the fitted parameters for the different replicates in the
unknown sample runs points out an obvious flaw with this
parameterization. There is clearly more than one parameter set with
an equally good fit to the data, meaning hidden dependencies in the
parameters that simple rescaling cannot uncover. Determining these
dependencies through alternative rescalings and singular
perturbation analysis, and through other parameter estimation
techniques is the subject of ongoing research.

\begin{figure}
\centerline{\epsfig{file=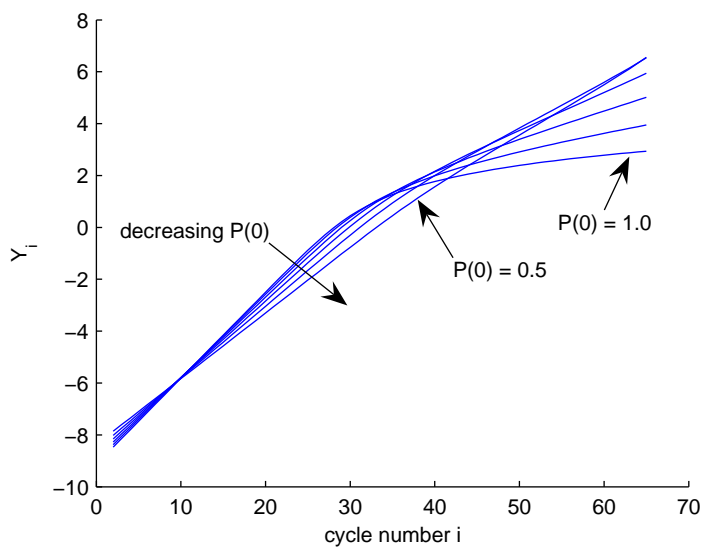}}
\caption{Integration of the two stage model with Taq Dynamics, with varying initial primer amount,
$p_0(0)$.}
\label{p0figure}
\end{figure}

Turning to the problem of realistic initial primer concentration, we
found that for values much less than 1.0 the model did not capture
the non-logistic behavior of the data. This can be seen in figure
\ref{fittedodemodel} b), where the curves have a concave down
portion near the end of the run, indicating the slowing of growth of
the amplicon. In figure \ref{p0figure} we illustrate this by
graphing the logarithmic regression variable for differing values of
initial primer, with all other parameters fixed.   In performing
parameter estimation, and taking lower initial primer concentration,
we found that we could not overcome the logistic-type behavior by
varying other reaction parameters.   A heuristic explanation for
this is that the nucleotides must be running out as well as primer
in this model to get the slower than logistic growth at the end of
the run. The equation for $\dot{n}$ (from the two dimension
reduction) is
\[
\dot{n}=-\beta (q(0)-q) n,
\]
so the rate of loss of $n$ is proportional to $n$.
The amount of dsDNA created is equal to $n(0)-n(t)$ at the end of
the cycle, so slower loss of $n$ means slower growth of $d$ and hence $s$.
 If the initial primer concentration is not close to
that of the initial nucleotide concentration  the quantity $n$ (which is
depleted necessarily at the same rate as $p$) will not become small,
and the rate of creation of $s$ will not slow accordingly.

\section{Discussion and Conclusions}
In this paper we have analyzed a sequence of models for the
reactions of PCR.  Exponential growth, the first order approximation
of simple doubling of the DNA strands, is replaced by a logistic
model which captures the sigmoidal nature of the amplification
curves.   Both these models are in common use in current devices. We
postulated a variation on the logistic model where the efficiency
decreases in time, and were able to fit the data with good results.
This, however, is less satisfactory than a model built directly from
the reactions that captures the decrease in efficiency as cycle
number increases.  We then built two such models, one that does not
include the enzyme Taq directly, and a second which does.

The model that did not include Taq dynamics is solvable
analytically, at least for each stage.  A map is created by linking
the closed form solutions through their initial conditions.   It was
found that the data could be well estimated by reasonable parameter
values if the initial amount of primer was taken close to that of
the initial amount of nucleotide. This is not the protocol followed
in the experiment, however, where for a 200 BP sequence the
primer:nucleotide ratio is about 1:4.   It appears that the amount
of nucleotides in the reaction must decrease significantly by the
end of the reaction in order to obtain a decrease in overall
reaction efficiency.  For this to happen the initial concentration
of primer must about that of the nucleotide.   From these
observations we concluded that the model without Taq dynamics did
not capture the full behavior of the amplification curves.

The second model is built on reactions that include the formation of
a complex of primed ssDNA and the Taq enzyme. The equations of stage
I could still be solved analytically, but this was not possible for
stage II.  Instead limiting cases of long integration time in stage
I or stage II or both were considered, and by analyzing the
amplification curves created by these maps we concluded that none of
the limiting cases, including variable annealing and long time
extension, would create the desired behavior at the end of the
reaction.   If the extension stage of the reaction was very fast
compared to the annealing phase, you might expect to capture the
qualitative behavior with this last case.

As none of the limiting cases created the qualitative
end-of-reaction behavior we thought the data demonstrate so clearly,
we proceeded to parameterize the full equations for the two stage
model with Taq dynamics.  With a map created from solutions to these
we were able to find parameters that captured the decreased
efficiency at the end of the reaction, but only if initial primer
concentration was again roughly the same size as the initial
concentration of nucleotides.   The extension phase of the reaction
would need to slow accordingly to fit this aspect of the behavior.
Also, multiple sets of parameters were found to fit the same
amplification run, indicating hidden dependencies in the parameters
that simple rescaling does not uncover.

While we were not able to completely explain the reduced efficiency
seen in the data with our suite of models, we were able to determine
what portions of the model were important in capturing its non-logistic
character.    Competing reactions at higher cycle numbers most
certainly will have an effect on the efficiency, especially with the
lower initial copy number runs.
Future work will include analyzing these dependencies both
numerically and analytically, and using the two stage model to
seek reaction protocols that minimize time to almost complete
creation of amplicon, and maximize yield for a fixed total cycle
time.

\section*{Acknowledgements}
We would like to thank Dr. David Eyre, Idaho Technology, for
suggesting the problem and for providing us with IT data and much
consultation. This work was partially supported by a Utah State
University Community/University Initiative grant, a USU Center for
Integrated Biosystems seed grant, and one undergraduate (John
Goldes) was supported by NSF-EPSCoR funds at the University of
Montana-Missoula.

\section*{Appendix:  Linear stability analysis for stage II with Taq Dynamics}

This stability analysis uses the two dimensional system that takes advantage of the conserved quantities,
 $s'+c+d=s'(0)$, $q+c=q(0)$, and $n+d=1$.   In the following
 $\gamma,\beta$ and all the initial quantities are positive.
\[
\dot{q}=-\gamma(s'(0)-q(0)-1+q+n)q+q(0)-q+\beta(q(0)-q)n
\]
\[
\dot{n}=-\beta(q(0)-q)n.
\]

\subsection*{The fixed points}
Solving $\dot{q}=0$ and $\dot{n}=0$ simultaneously gives three fixed points.
\[
f.p. 1=( \bar{q}=q(0), \bar{n}=1-s'(0))
\]
\[
f.p. 2=(\bar{q}=\frac{1}{2\gamma}(\gamma(q(0)+1-s'(0))-1+ \sqrt{(-\gamma(q(0)+1-s'(0))+1)^2+4\gamma q(0)}),
\]
\[
\bar{n}=0)
\]

The third fixed point has
\[
\bar{q}=\frac{1}{2\gamma}(\gamma(q(0)+1-s'(0))-1- \sqrt{(-\gamma(q(0)+1-s'(0))+1)^2+4\gamma q(0)}),
\]
which is always negative.  Proof:

$\bar{q}$ is a root of the polynomial $\gamma q^2+(-\gamma(q(0)+1-s'(0))+1)q-q(0)$.
Let $b=-\gamma(q(0)+1-s'(0))+1$.
Then
\[
\sqrt{b^2+4\gamma q(0)}>b,
\]
therefore,
\[
-b-\sqrt{b^2+4\gamma q(0)}<0.
\]
Since f.p.3 is not of physical importance, we will only analyze the
stability of f.p. 1 and f.p. 2.

\subsection*{Stability of f.p.1}
The Jacobian for the reduced system is
\[
 \left[ \begin {array}{cc} -\gamma\,q-\gamma\, \left( s'(0)-q(0)-1+q+n
 \right) -1-\beta\,n&-\gamma\,q+\beta\, \left( q(0)-q \right)
\\\noalign{\medskip}\beta\,n&-\beta\, \left( q(0)-q \right) \end {array}
 \right]
\]

 Let $s'(0), q(0)>0,c(0)=0,n(0)=1$, and $d(0)=0$.   Also let $\gamma,\beta>0$.
 For f.p. 1, the Jacobian becomes
 \[
\left[ \begin {array}{cc} -\gamma\,q(0)-1-\beta\, \left( 1-s'(0) \right) &-
\gamma\,q(0)\\\noalign{\medskip}\beta\, \left( 1-s'(0) \right) &0\end {array}
 \right]
\]
The eigenvalues are,
\[
1/2(-\gamma q(0)-1-\beta(1-s'(0))\pm \sqrt{(\gamma q(0)+1+\beta(1-s'(0)))^2-4\gamma q(0)\beta(1-s'(0))}).
\]

The location of the first fixed point and the sign of the nonzero eigenvalues depend
on the relationship between $n(0)=1$ and $s'(0)$.

Case1: $1>s'(0)\Rightarrow 1-s'(0)>0$. Both coordinates of the first
fixed point are non-negative.  The eigenvalues are real
 and negative.
 They are real because the discriminant is greater than zero.
 Proof:
 \[
 (\gamma q(0)+1+\beta(1-s'(0)))^2-4\gamma q(0) \beta(1-s'(0))
 \]
 \[
 =\gamma^2 q(0)^2+2\gamma q(0)+2\gamma q(0) \beta(1-s'(0))+2\beta(1-s'(0))+1+
 \beta^2(1-s'(0))^2-4\gamma q(0) \beta(1-s'(0))
 \]
 \[
=\gamma^2 q(0)^2 -2\gamma q(0) \beta(1-s'(0))+\beta^2(1-s'(0))^2+2\gamma q(0)+2\beta(1-s'(0))+1
 \]
 \[
 =(\gamma q(0)-\beta(1-s'(0)))^2+2\gamma q(0)+2\beta(1-s'(0))+1>0.
 \]
 They are both negative.  Proof:

 Let $b=\gamma q(0)+1+\beta(1-s'(0)).$  Then $b^2>4\gamma q(0) \beta(1-s'(0))$ as shown above.
 \[
 \sqrt{b^2-4\gamma q(0) \beta(1-s'(0))}<b,
 \]
 therefore,
 \[
 \frac{-b+\sqrt{b_2-4\gamma q(0) \beta(1-s'(0))}}{2}
 \]
 and
 \[
 \frac{-b-\sqrt{b_2-4\gamma q(0) \beta(1-s'(0))}}{2}
 \]
 are both $<0.$

 Case 2: $n(0)=1=s'(0)\Rightarrow 1-s'(0)=0.$
 The first fixed point becomes $(q(0),0)$ which still is nonnegative.
 The eigenvalues become:
 \[
\frac{1}{2}(-\gamma q(0)-1+\sqrt{(\gamma q(0)+1)^2})=\frac{1}{2}(-\gamma q(0)-1+\gamma q(0)+1)=0
 \]
and
 \[
 \frac{1}{2}(-\gamma q(0)-1-\sqrt{(\gamma q(0)+1)^2})=\frac{1}{2}(-2\gamma q(0)-2)=-\gamma q(0)-1.
 \]
 For this case, there is only one nonzero eigenvalue, $-\gamma q(0)-1$, which is negative.

 Case 3:  $n(0)=1<s'(0)\Rightarrow 1-s'(0)<0.$
The coordinate, $\bar{n}=1-s'(0)$, of the first fixed point is now
negative.  The eigenvalues,
 \[
 1/2(-\gamma q(0)-1-\beta(1-s'(0))\pm \sqrt{(\gamma q(0)+1+\beta(1-s'(0)))^2-4\gamma q(0)\beta(1-s'(0))}),
 \]
are real.
 Proof:

 Let $\alpha>0$ and let $1-s'(0)=-\alpha.$ Let
\[
 b=\gamma q(0)+1+\beta(-\alpha)=\gamma q(0)+1-\beta\alpha.
 \]
\[
b^2-4\gamma q(0)\beta(-\alpha)=b^2+4\gamma q(0)\beta\alpha,
\]
therefore
\[
b^2+4\gamma q(0)\beta\alpha>0.
\]
One of these eigenvalues is positive and the other is negative.
Proof:

$b^2+4\gamma q(0)\beta\alpha>0,$ so $\sqrt{b^2+4\gamma q(0)\beta\alpha}>b.$  Then
\[
\frac{-b+\sqrt{b^2+4\gamma q(0)\beta\alpha}}{2}>0
\]
 and
\[
\frac{-b-\sqrt{b^2+4\gamma q(0)\beta\alpha}}{2}<0.
\]

\subsection*{Stability of f.p.2}
For f.p.2, the Jacobian becomes
\[
\left[ \begin {array}{cc} -\gamma\,\bar{q}-\gamma\, \left( s'(0)-q(0)-1+\bar{q} \right)
-1&-\gamma\,\bar{q}+\beta\, \left( q(0)-\bar{q} \right) \\\noalign{\medskip}0&-\beta
\, \left( q(0)-\bar{q} \right) \end {array} \right]
\]
where $\bar{q}=\frac{1}{2\gamma}(\gamma(q(0)+1-s'(0))-1+ \sqrt{(-\gamma(q(0)+1-s'(0))+1)^2+4\gamma q(0)}).$
The eigenvalues are
\[
\frac{\beta}{2\gamma}(-\gamma q(0)+\gamma(1-s'(0))-1+\sqrt{(\gamma q(0)-\gamma(1-s'(0))+1)^2+4\gamma^2 q(0)(1-s'(0))},
\]
\[
- \sqrt{(\gamma q(0)-\gamma(1-s'(0))+1)^2+4\gamma^2 q(0)(1-s'(0))}.
\]

Case1:  $1-s'(0)>0$
For this case $\bar{q}$ is positive.  Proof:

Let $b=-\gamma(q(0)+1-s'(0))+1.$ $b^2+4\gamma q(0)>0$, so $\sqrt{b^2+4\gamma q(0)}>b.$
Therefore,
\[
\frac{-b+\sqrt{b^2+4\gamma q(0))}}{2\gamma}>0.
\]
The eigenvalues are both real.  Proof:

\[
(\gamma q(0)-\gamma(1-s'(0))+1)^2>0.
\]
Also, for $1-s'(0)>0,$ $4\gamma^2 q(0)(1-s'(0))>0.$  Therefore, the discriminant is positive.  The eigenvalue
\[
- \sqrt{(\gamma q(0)-\gamma(1-s'(0))+1)^2+4\gamma^2 q(0)(1-s'(0))}
\]
is clearly negative. The other eigenvalue is positive for this case.  Proof:

Let $b=\gamma q(0)-\gamma(1-s'(0))+1.$
\[
b<\sqrt{b^2+4\gamma^2 q(0)(1-s'(0))}
\]
therefore,
\[
-b+\sqrt{b^2+4\gamma^2q(0)(1-s'(0))}>0
\]
Case 2: $1-s'(0)=0.$
For this case, $f.p.1=f.p.2.$ Proof:

\[
\bar{q}=\frac{1}{2\gamma}(\gamma q(0)-1+\sqrt{(-\gamma
q(0)+1)^2+4\gamma q(0)})
\]
\[
=\frac{1}{2\gamma}(\gamma q(0)-1+\sqrt{-\gamma^2 q(0)^2-2\gamma
q(0)+1+4\gamma q(0)})
\]
\[
=\frac{1}{2\gamma}(\gamma q(0)-1+\sqrt{-\gamma^2 q(0)^2+2\gamma q(0)+1}
\]
\[
=\frac{1}{2\gamma}(\gamma q(0)-1+\sqrt{(\gamma q(0)+1)^2}
\]
\[
=\frac{1}{2\gamma}(\gamma q(0)-1+\gamma q(0)+1)=\frac{1}{2\gamma}(2\gamma q(0))
\]
\[
=q(0).
\]
$f.p.2$ becomes $\bar{q}=(q(0),\bar{n}=0)$ which equals $f.p.1.$

Case 3:  $1-s'(0)<0.$  The eigenvalues remain real for this case, but the one that was positive in Case
 1 becomes negative.  Proof:

Let $-\alpha=1-s'(0),$ $\alpha>0,$ and let $b=\gamma q(0)+\gamma\alpha+1.$  Then
\[
b>\sqrt{b^2-4\gamma^2 q(0)\alpha}
\]
therefore,
\[
-b+\sqrt{b^2-4\gamma^2q(0)\alpha}<0
\]

\subsection*{Summary}
For Case 1, f.p.1 is a sink, but becomes a saddle as s'(0) becomes larger than n(0)(for Case 3).
The opposite occurs for f.p.2.  It starts out as a saddle (for Case 1) and becomes a sink (for Case 3).
This indicates a transcritical bifurcation when the initial
amount of nucleotides equals the initial amount of primed single-strand DNA.

\end{document}